# Nonlocal metasurfaces for spectrally decoupled wavefront manipulation and eye tracking

Jung-Hwan Song, Jorik van de Groep, Soo Jin Kim, and Mark L. Brongersma[*]

Geballe Laboratory for Advanced Materials, Stanford University, Stanford, CA 94305, USA.

[*]brongersma@stanford.edu

**Abstract**

Metasurface-based optical elements typically manipulate light waves by imparting space-variant changes in the amplitude and phase with a dense array of scattering nanostructures. The highly-localized and low optical-quality-factor ($Q$) modes of nanostructures are beneficial for wavefront-shaping as they afford quasi-local control over the electromagnetic fields. However, many emerging imaging, sensing, communication, display, and non-linear optics applications instead require flat, high-$Q$ optical elements that provide notable energy storage and a much higher degree of spectral control over the wavefront. Here, we demonstrate high-$Q$, nonlocal metasurfaces with atomically-thin metasurface elements that offer notably enhanced light-matter interaction and fully-decoupled optical functions at different wavelengths. We illustrate a possible use of such a flat optic in eye tracking for eye-wear. Here, a metasurface patterned on a regular pair of eye-glasses provides an unperturbed view of the world across the visible spectrum and redirects near-infrared light to a camera to allow imaging of the eye.

**Main text**

The emergence of new technologies can be a tremendous driver for innovation. We currently witness this with the rapid development of novel handheld and wearable technologies. They are home to a microcosm of electronic and optical devices that have to efficiently work together to perform a complex set of display, sensing, imaging, filtering, computation, and illumination functions. The need for such devices to be lightweight, fashionable, and operate at high speed and low power has prompted some of the most thought-provoking design challenges[1–5]. Nanostructured layers termed metasurfaces[6,7] have presented many elegant, new solutions for the optical components by offering a compact form-factor, multi-functionality[8,9], very high-numerical apertures[10,11], minimal aberrations[12–14], and control over the light field[15,16]. Now, in the development of augmented and virtual reality (AR/VR) systems new functionalities are required for which even metasurface-based flat optics does not appear to have solutions. Functions that require very high spectral and angular control have proven particularly hard to achieve. These are not afforded by conventional metasurfaces that have largely relied on engineering of a local optical response, where single or small groups of closely-spaced nanostructures are designed to imprint independent and spatially-variant phases onto incident light waves. As the optical quality factor of subwavelength nano-scatterers tends to be low, they offer similar optical functions for light waves across broad ranges of wavelengths and incident angles. The realization of narrow band responses in planar structures has typically been the domain of photonic crystals[17] and guided-mode resonator (GMR) devices[18–23] that display nonlocal responses involving optical modes that extend over many elements. Recent works on nonlocal metasurfaces have employed the concepts of GMRs[24] and bound states in the continuum (BIC)[25–28] to realize high-$Q$ flat optics capable of selectively manipulating light at certain wavelengths and incident angles at which nonlocal modes can be excited. One major limitation of these high-$Q$, non-local metasurfaces has been that their optical functions at different wavelengths are strongly tied together through their band structure and spectrally-decoupled functions appear impossible to realize. In this work, we demonstrate that the introduction of spectrally-dependent materials absorption can fully-decouple optical functions at different wavelengths. We start by showing how the need for this capability arose in an attempt to develop creative, new solutions for eye-tracking (ET) technology in wearables and AR devices.

ET has a myriad of applications in medicine, psychology and neuroscience, marketing research, sports, and gaming[29–32]. Optical ET techniques have successfully been commercialized and are integrated into regular and AR/VR eyewear [supplementary information table S1]. Figure 1a shows a photograph of a basic ET prototype that we constructed with a large area (4 cm$^2$) high-$Q$, nonlocal metasurface created on the eye-facing surface of a pair of glasses. A NIR-LED emitting at 870 nm illuminates an artificial eye and the metasurface redirects the scattered light from the eye toward a miniature camera that is attached to one of the arms. The metasurface should provide the person wearing the glasses with an unperturbed view of the outside world, while a high-speed camera can image and follow the eye's motion. In an existing approach, two or more cameras are attached to the frame in such a way that they do not block the user's view and allow capture of side-view images of the eye. To assess the gazing direction from side-view images unfortunately requires an undesirably large amount of computational processing power/time and the use of complex imaging algorithms. For this reason, it is ideal to acquire a front-view image with a single camera. This can also be accomplished with other compact optical elements (e.g. gratings, miniature prisms, local metasurfaces, and waveguides) placed directly in front of the eye, but these dispersive components inevitably produce unwanted rainbows. This is very distracting when used in environments with bright optical sources, such as the sun[33,34]. There are currently no compact ET systems that offer the required high transparency (> 80%) across the visible, acceptable efficiencies (> 10%) for redirecting near-IR photons for imaging purposes, and strong suppression of rainbows (> 3 orders of magnitude)[3,12,33,35–38].

It is clear that a new type of optical element needs to be created that is capable of performing different, independent operations on light waves at different frequencies and incident angles. In our example, all light waves in the visible (VIS) spectrum should be allowed to pass completely unperturbed, independent of the incident angle. However, in the NIR we aim to redirect light across a narrow band of wavelengths and a small range of angles to a camera for imaging. At least in principle, linear optical elements are well-suited for the task of sorting out photons in overlapping beams by their fundamental characteristics (wavelength, polarization state or propagation direction). Physically, it is a matter of manipulating wave interference effects and the relevant mathematics is comprised of simple additions and subtractions[39]. However, in practice it can be extremely challenging to design a thin, ideally single-layer optical element that acts very differently on different types of waves. Only recently, new physics concepts[40] and inverse design for nanophotonics[41] have started to enable such tasks. Given the tremendous design flexibility for metasurfaces with millions of scattering nanostructures, this technology would appear promising. However, a clever arrangement of the nanostructures turns out to be insufficient to solve this challenge and we will argue that it is critical to achieve a high-$Q$, nonlocal response by engineering a long-range, coherent coupling between many designer elements in the metasurface.

To see the benefits of our metasurface design, we first consider the properties of a conventional grating comprised of 30-nm-thick Cr strips whose dimensions where optimized to redirect 10% of a normally-incident NIR beam through the first diffracted order. The view through such a grating is plagued by the presence of intense rainbows (see Fig. 1b) that are a direct consequence of the local scattering response of the Cr strips. Figures 1c, d show the same scene as viewed through two of the newly-proposed high-$Q$, nonlocal GMR metasurfaces. They are comprised of either 7-nm-thick or 3-nm-thick poly-crystalline Si (pSi) strips placed on a $Si_3N_4$ dielectric waveguide (Fig. 1e). These structures effectively suppress the rainbows and nanometer-scale changes in the thickness of the grating elements strongly impact the intensity of the rainbows. To understand the difference in behavior from regular Cr gratings and the high-$Q$ GMR metasurface, it is of value to go back to the early modeling work on GMR structures by Hessel and Oliner[18]. They analyzed how surface-relief gratings etched into waveguides can couple free-space waves to quasi-guided-modes at selected resonant frequencies. The amplitude of the grating elements determines the coupling efficiency in/out of the waveguide and thus also the radiative $Q$ for the resonances. The period of the GMR gratings has traditionally been chosen sufficiently small to suppress the generation of diffracted-orders in the super- and substrates. In this configuration, GMRs have been used to couple light into waveguides and cavities and as high-finesse optical bandpass filters[19–21]. For our application, we create GMRs with larger periods and redesign them so that they can function as metasurface-based flat optics capable of steering light at a preselected wavelength of NIR light (in our example 870 nm) while allowing an unperturbed transmission across the visible without the appearance of rainbows.

The optical spectra in Figs. 1f,g quantitatively compare the distinct optical behavior of the Cr grating (Fig. 1b) and optimized high-$Q$ GMR metasurface (Fig. 1d). The Cr grating displays a low transmissivity across the visible, which is linked to strong absorption and scattering by the metal structures. We can see undesired diffraction across the entire visible range, consistent with Fig. 1b. This results from diffraction into second- and higher-order diffracted beams. In contrast, the 3-nm-thick GMR metasurface shows a very high transmissivity at normal incidence across the visible range (89% on average). This can be expected as the grating is at least 2 orders of magnitude thinner than the absorption depth in pSi. Moreover, the $Si_3N_4$ waveguide and silica cover also serve as an optimized, double-layer antireflection coating [supplementary information S2]. The transmission spectrum for the GMR metasurface shows a sharp spectral feature near 870 nm that can be attributed to a GMR. This resonance provides a second, indirect pathway for the light through the structure. At the resonant wavelength, about 40% of the light is taken out of the incident beam and redirected into four first-order diffracted beams (+1 and -1 in forward and backward directions). Whereas the scattering by a single metasurface/scattering element is weak, a high diffraction efficiency nonetheless results from constructive interference of many coherent scattering elements that work together on resonance to first couple and then redirect the light. The diffraction efficiency spectrum (Fig. 1g) shows peaks with an asymmetric lineshape that come about from the interference of the light radiating from the GMR and the directly transmitted light (i.e. the background).

Based on the high diffraction efficiency seen at 870 nm, one would naturally expect to see high diffraction-efficiencies in the visible range as well. However, the GMR resonances in the VIS (at 608 nm, 496 nm, and 470 nm) display a very low diffraction-efficiency (< 0.07%), explaining the virtual absence of rainbows. The difference in behavior in the VIS and NIR lies in the spectral absorption properties of the pSi metasurface elements, as explained next. The dashed curves in fig. 1f,g show the transmission and diffraction efficiency for the case that the materials absorption in Si is artificially set to zero. They very nicely highlight how the materials absorption in the ultrathin metasurface elements can control the flow of light through the device on resonance; In the presence/absence of materials absorption, the diffracted orders can selectively be turn off/on without notably impact the response off-resonance.

The flow of light through the GMR metasurface is controlled by the relative magnitudes of the scattering and absorption efficiencies of the metasurface elements. These can be engineered through a choice of the material and geometry for these elements. For our metasurface, we obtain very different behavior in the VIS and NIR by using pSi elements. The absorption depth of pSi changes by several orders of magnitude between these spectral ranges (Fig. 1g). Figure 2a illustrates the power flow for incident planewaves in the VIS and NIR. In both spectral ranges, light can couple to the GMR by picking up the second-order grating momentum ($=4\pi/p$), where $p$ is the grating period [supplementary information S3]. The dispersion diagram in Fig. 2b shows this process for 870 nm transverse electric (TE) polarized light. Subsequent radiative decay from the GMR into the first diffracted-order then produces the desired steered beam for imaging. The decoupling by the 3-nm-thick grating is very weak and this enables a resonance with a high $Q$ of 8700 at $\lambda = 870$ nm. This $Q$ is radiation-limited as the pSi shows weak material's absorption at this wavelength. As such, the light is trapped inside the $Si_3N_4$ slab waveguide for many optical cycles ($=Q/2\pi$) and a strong buildup of the internal fields in the GMR ($|\mathbf{E}|^2/|\mathbf{E}_{inc}|^2 \approx 2500$) is observed. In turn, the large internal fields enable high diffraction efficiencies of 11.1% for the first-order reflected beams, consistent with temporal coupled mode theory (CMT) [supplementary information S4]. The diffraction is clearly noticeable from the checkerboard pattern of nodes in the electric field intensity profile shown in Fig. 2c. In contrast, in the visible part of the spectrum the pSi material's absorption is significantly stronger. As a result, the quality factor is absorption-dominated and the field buildup in the GMR results in a strongly enhanced light absorption. In this wavelength range, the energy in the GMR dissipates before it decouples to free-space. Physically, the enhanced absorption results from a long interaction length with the absorbing pSi grating elements as the light is guided along the nitride waveguide. This results in a low first-order diffraction efficiency (< 0.1%), as confirmed by CMT and a power flow analysis [supplementary information S4]. For the directly transmitted light there is no noticeable energy storage and we find that the essentially single-pass transmission through the 3-nm-thick pSi grating provides a very high (close to 90%) transmittance [supplementary information S5]. This shows we can fully deactivate the diffractive optical functions in the visible spectrum simply by introducing some materials absorption in the atomically-thin metasurface elements. At this point it is worth restressing that the Cr strips in the conventional grating only support low-$Q$, local scattering resonances. Such resonances do not afford the significant field buildup that is required to achieve independent control over the power flow paths in the VIS and NIR.

There are several trade-offs in the metasurface design. Thicker pSi gratings elements can more effectively scatter than absorb light and can thus offer more intense first-order diffracted beams. Figure 2d shows the simulated dependence of the diffraction efficiency for the TE-polarized GMR near 870 nm into the first-order reflected beam used for ET. Here, we take a representative refractive index of pSi (=3.7+0.005$i$) in this spectral regime[42]. Despite the very weak material's absorption of pSi at this wavelength, the absorption ends up dominating the total quality factor ($1/Q = 1/Q_A+1/Q_R$) for very thin strips (~1 nm) as a result of the large optical path length inside the GMR. The peak diffraction efficiency then increases as a function of the grating height and saturates as the scattering by the strips starts dominating the absorption [supplementary information S6]. At a 2 nm grating height, the diffraction efficiency already exceeds 4%. Figure 2e summarizes the diffraction efficiency into the unwanted, first-order transmitted beam for the TE-polarized GMR in the visible (near 605 nm). In this spectral region, the pSi is significantly more absorptive ($n = 4.0 + 0.1i$) and the peak diffraction efficiency is decreased by 2 orders of magnitude as compared to the NIR. The peak diffraction efficiency falls below 0.1% for grating heights smaller than 4 nm, which we therefore set as the maximally acceptable grating height. Combined, these requirements locate the optimum grating height around 3 nm. Finally, such a grating height is also thin enough to fulfill the >80% transmittance requirement (see Fig. 2f).

The analysis above indicates that we need atomic-scale control over the Si grating thickness across the 4 cm$^2$ area of the metasurface. The scanning electron microscopy (SEM) image in Fig. 3a shows how we can achieve such high accuracy by adopting standard photolithography and Si-compatible processing [supplementary information S7]. A pSi deposition followed by a slow thermal Si oxidation process enables sub-nanometer control over the pSi layer height [supplementary information S8]. The cross-section of the grating pattern shows a 30-nm-deep over-etching of the pSi layer into the Si$_3$N$_4$ slab (Fig. 3b). The over-etching increases the radiative decay rate compared to that for the non-over-etched case [supplementary information S9]. Cross-polarized transmission measurements remove the direct, non-resonant transmission channel for the metasurface and clearly show the sharp resonance peaks from the GMRs (Fig. 3c). The measured quality factor ($Q = 540$) is lower than in the simulation ($Q = 1400$), which we attribute to the increased radiative decay rate by fabrication imperfections.

Using a home-built angle-resolved confocal spectroscopy setup, we characterize the optical dispersion (Fig. 3d) as well as the power efficiency of the resonant diffraction (Fig. 3e) by the high-$Q$ GMR metasurface [supplementary information S10]. Figure 3e shows the measured peak diffraction efficiency spectra for both TE and TM polarizations. The measured data agrees well with those of the corresponding simulations [supplementary information S5]. We find that the TE GMR at 864 nm delivers a 12.8% diffraction efficiency. The slightly larger peak diffraction efficiency compared to the simulation (11.1%) is attributed to a small deviation in the width of the pSi grating elements from the original design. This mode provides an excellent platform to redirect normally-incident NIR light from the eyes into the desired direction over a 60° angle, where the ET camera is positioned. The diffraction in the visible spectrum is suppressed to 0.07% on average while maintaining a high degree of zeroth-order transmission (> 85%) [supplementary information S5]. It is worth noting that the ability to diffract light at long wavelength and let it pass unperturbed at short wavelengths goes against one of the most basic traits of conventional, periodic gratings.

As a final step, we demonstrate the possible use of the proposed metasurface in a basic, prototype ET system. Figure 4a shows a front-view of the prototype ET glasses where the location of the metasurface is outlined. This front-view of the eye is also seen by the camera after a redirection by the metasurface. We develop a basic ray-tracing approach to describe the image formation by the metasurface [supplementary information S11]. It is used to transform a collection of points in the object space to a set of stigmatic points in a virtual image as depicted in Fig. 4b. The stigmatic points can be located by considering the path of two rays originating from a single point on the eye (e.g. *f*). The diffracted rays can be extended back from the metasurface to find the stigmatic points (e.g. *f'*). From such an analysis, it is clear how the camera can capture the front-view of the eye through a diffracted order. It further illustrates how the eye's image is compressed along the *z*-direction and pulled towards the metasurface. This action enables us to track large lateral motions of the pupil and spatially resolve features of the eye, no matter how far the eye is located from the metasurface [supplementary information S12]. Possible concerns with imaging through a diffractive order are the achievable image resolution and the impact of image distortion[43,44] along the *x*-direction. The high-$Q$ of the GMR metasurface limits the range of angles that can effectively couple to the guided mode and thus can be used for imaging. A detailed analysis [supplementary information S13] shows that the image resolution is limited by the numerical aperture of the system and given as $d = \lambda_r/2\text{NA} = \lambda_r Q/2\pi n_g$, where $\lambda_r$ and $n_g$ are the wavelength and group index of the relevant guided mode. Plugging in the experimentally obtained values of $\lambda_r$ (= 864 nm), $Q$ (= 540), and $n_g$ (2.04) we find an image resolution of ~40 μm. We chose our imaging optics to closely match the horizontal image resolution of the high-$Q$ GMR metasurface and an image taken from a United States Air Force (USAF) resolution target confirms our ability to easily resolve 0.63-mm-sized horizontal and vertical bars. This image resolution is also sufficient to image the key features of the eye, even at the large rotation angles shown in Fig. 4d. Our experimental investigation of the eye-tracking performance demonstrates that the measured gaze directions based on the diffracted images are in excellent agreement with the actual gaze directions (to within ~1° deviation), consistent with what is generally required from commercial eye tracking systems [supplementary information S15]. If needed, a higher image resolution can be achieved using a lower $Q$ metasurface at the cost of increased rainbow artifacts. The bandwidth of the source should be matched to the GMR resonance to increase the overall power-efficiency. The diffraction efficiency into a desired order can also be enhanced further by tailoring the metasurface building blocks. We numerically demonstrate that asymmetric diffraction is possible with an efficiency of 53% of the incident planewave redirected into the diffractive channel used for imaging [supplementary information S16].

In this work, we have shown how high-$Q$, nonlocal metasurfaces can be created that offer independent functions across different wavelength bands. Light at a preselected resonance wavelength can be stored in the high-$Q$ GMR metasurface to notably boost the internal fields ($|\mathbf{E}|^2/|\mathbf{E}_{inc}|^2 \approx 2500$) and light-matter interaction before it is released into a desired direction. A judicious choice for the absorbing materials used to create the metasurface building blocks can facilitate effectively control over the flow of light through the metasurface and allow different, uncoupled functions at different wavelengths. Since only the resonant wavelengths are manipulated, these structures can conveniently be stacked [supplementary information S17]. We expect these nonlocal metasurfaces to open many new functions for optical imaging, sensing, communication, display, and non-linear optics that complement those of conventional, low-$Q$ metasurfaces. The presented aspects of nonlocal metasurfaces also nicely complement the development of nonlocal metasurfaces for dynamic wavefront shaping[28] and analog optical computing.[45–47]

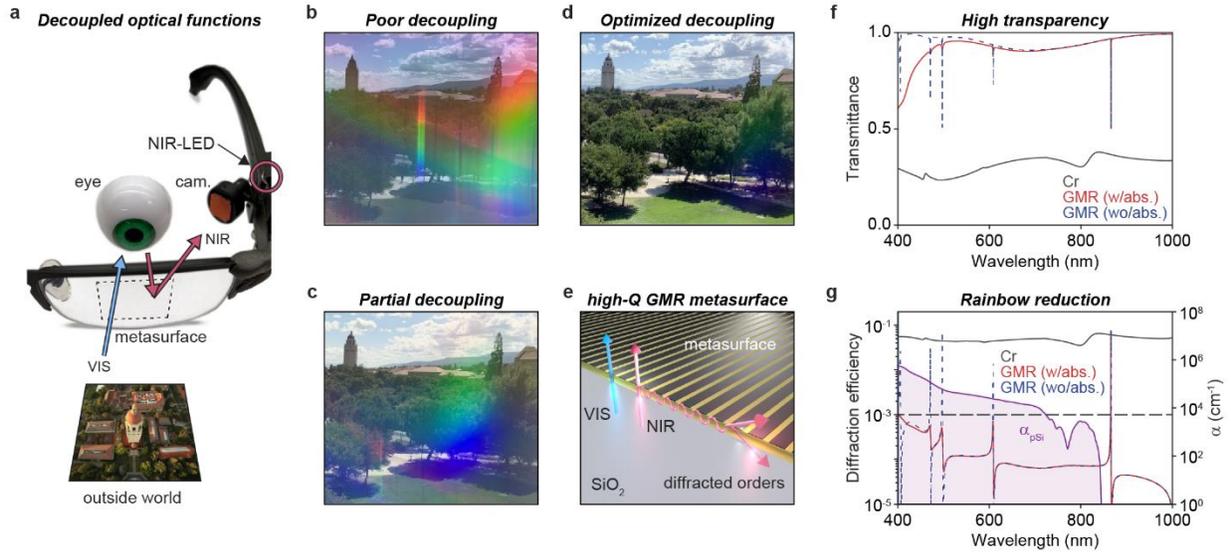

**Fig. 1 | A high-*Q* GMR metasurface facilitates decoupled optical functions at different wavelengths and optical eye-tracking without rainbows. a,** Photograph of our prototype ET glasses shows the importance of decoupling of optical functions in the NIR and the VIS spectral regions. Light from a NIR-LED scattered by the eye needs to be redirected to a camera by a judiciously patterned surface on the glass. At the same time, the pattern should not perturb the direct transmission of the light in the VIS region in order to allow an unperturbed view of the outside world (Hoover tower at Stanford University). **b,** Photograph taken through a glass surface with a 200-nm-thick $Si_3N_4$ anti-reflection coating patterned with a 30-nm-thick Cr grating on top showing multiple strong rainbows. Period and width of the Cr grating are 1 μm and 580 nm, respectively. **c,** Photograph taken through a GMR metasurface with a 7-nm-thick pSi grating showing a weak rainbow. GMR metasurface is comprised of 100-nm-thick $SiO_2$, 7-nm-thick pSi grating, 200-nm-thick $Si_3N_4$ slab, and glass substrate from top to bottom. Period and width of pSi grating are 1 μm and 900 nm, respectively. **d,** Photograph taken through a GMR metasurface with a 3-nm-thick pSi grating showing a strongly suppressed rainbow and a high, color balanced transmission. **e,** Schematic illustrating how a high-*Q* GMR metasurface affords unimpeded transmission of light across the VIS and narrow-band redirection of NIR light at 870 nm into a +1 diffracted order. **f,** Simulated zeroth-order transmittance spectra of the Cr grating and optimized GMR metasurface under normally-incident planewave illumination. The dashed blue curve shows the situation where the materials absorption in the Si is artificially set to zero. The polarization direction is in parallel to the grooves of the grating (TE). **g,** Simulated diffraction efficiency into the +1 diffracted-order transmission for the 30-nm-thick Cr grating and the optimized GMR metasurface under normally-incident TE polarized planewave incidence. Both structures reach ~10% efficiency at the target wavelength of 870 nm. The dashed blue curve shows the situation where the materials absorption in the Si is artificially set to zero. The horizontal dashed line indicates a 0.1% upper limit for visible light diffraction to avoid perceptible rainbows. The purple shaded region shows the absorption coefficient of grown pSi.

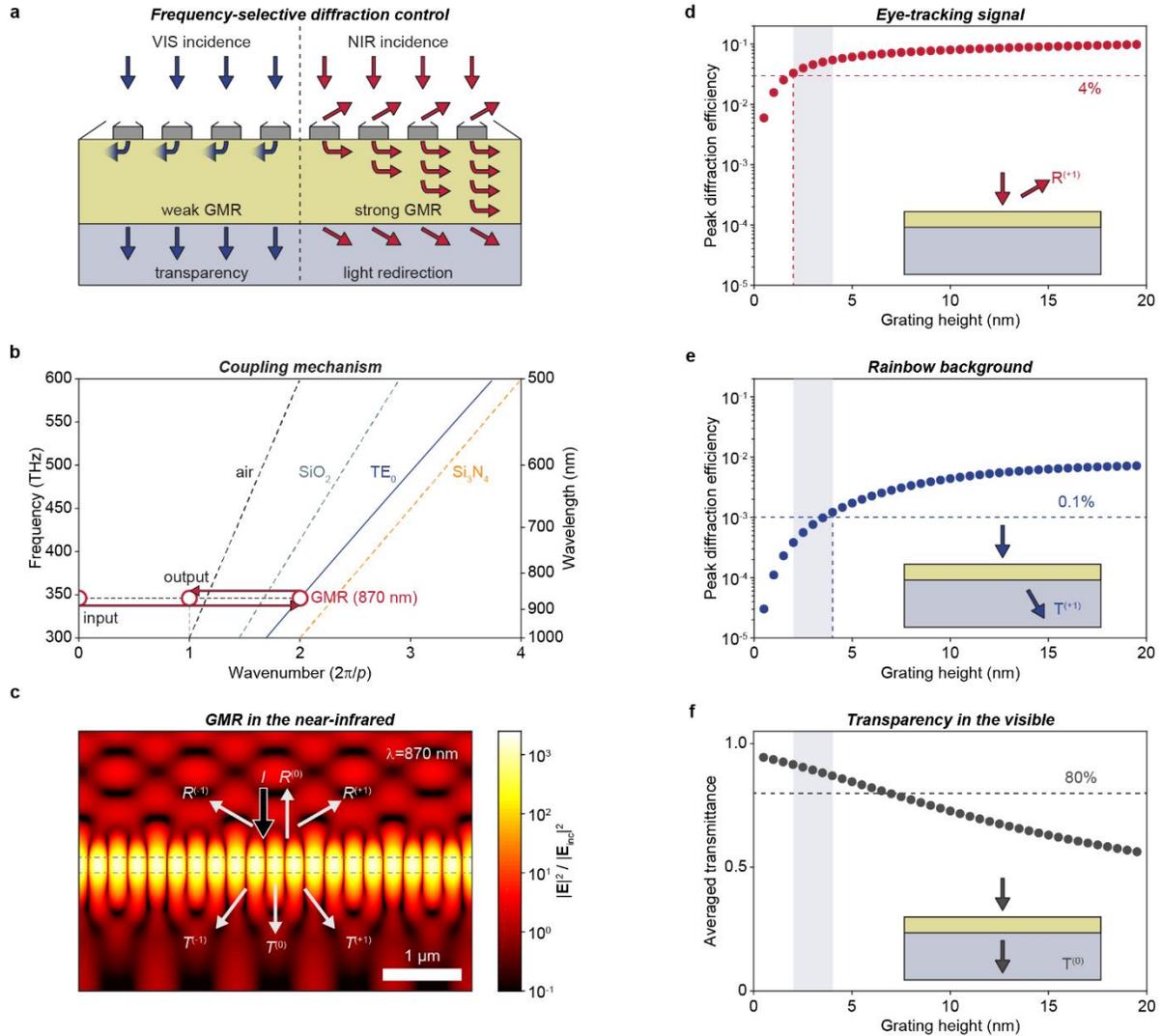

**Fig. 2 | Spectrally-selective resonant diffraction mechanism of GMR metasurface. a,** Schematic showing the optical power flow of input planewaves in the VIS and NIR into the transmitted and diffracted channels as mediated by the high-$Q$ GMR metasurface. **b,** Dispersion relation for the $TE_0$ mode supported by the $Si_3N_4$ waveguide. The black, cyan, and yellow dashed curves are the light lines for air, $SiO_2$, and $Si_3N_4$, respectively. The horizontal red arrow to the right indicates the excitation of the GMR by a normally-incident planewave. The red arrow to the left designates the decoupling from the $TE_0$ waveguide mode into the 1$^{st}$ diffracted order at grazing exit (i.e. close to the air light line). **c,** Electric field intensity profile of the TE GMR excited on resonance at 870 nm. The black arrow indicates the incident planewave and the 6 white arrows represent the output diffractive orders (including zeroth-orders). **d,** Resonant peak diffraction efficiency at the TE GMR in the NIR as a function of the Si grating height. All structural parameters except the grating height are fixed. **e,** Resonant peak diffraction efficiency of TE GMR in the VIS. **f,** Averaged transmittance of the GMR structure across the visible range (400 - 700 nm) for TE polarization. The dashed horizontal line shows a minimum required transmittance for a practical device.

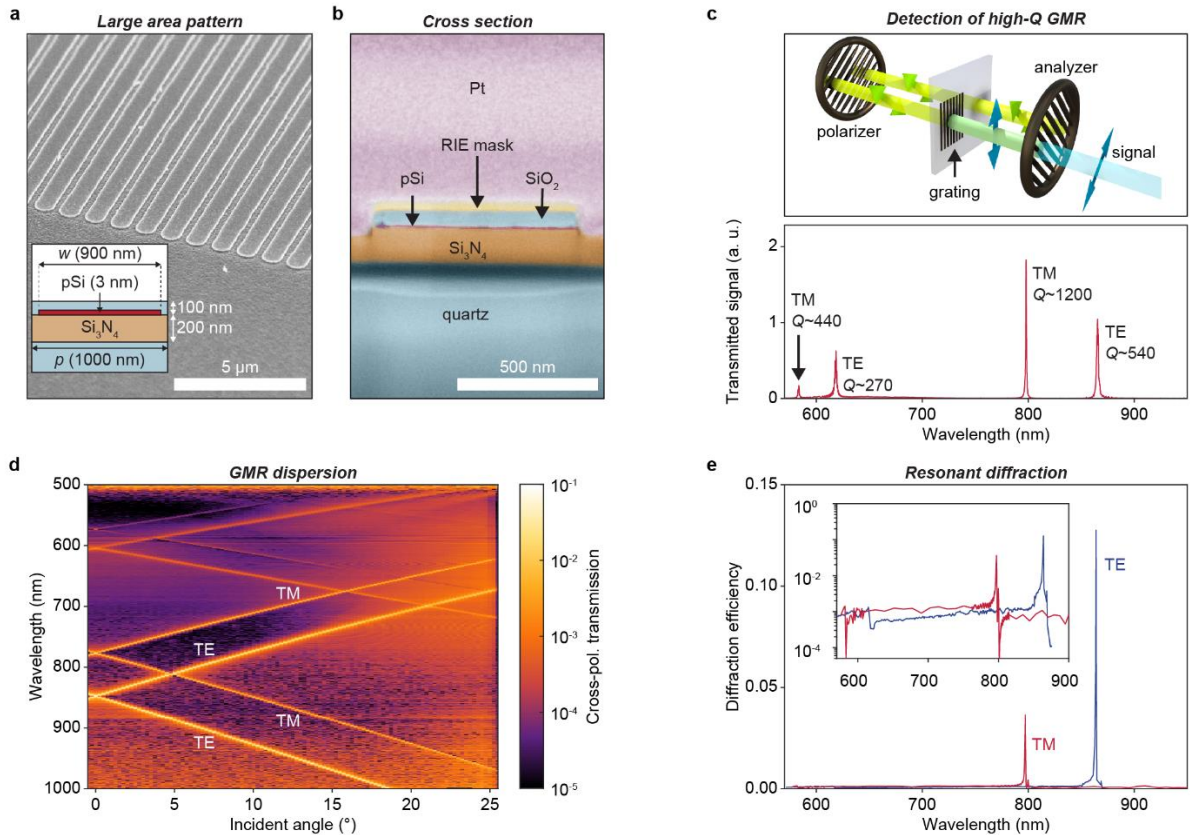

**Fig. 3 | Experimental realization of the spectrally-selective high-$Q$ GMR metasurface and spectrally-decoupled optical diffraction between NIR and VIS spectral regions. a,** SEM image of the fabricated high-$Q$ GMR metasurface after RIE etching. The following RIE mask removal and HSQ coating procedures complete the fabrication of the high-$Q$ GMR metasurface structure. The inset represents the designed cross section of a single unit cell. Superstrate and substrate are the air and quartz, respectively. The period ($p$) of the unit cell, the width ($w$) of pSi grating, and the thickness of $Si_3N_4$ slab are 1000 nm, 900 nm, and 200 nm, respectively. 100-nm-thick HSQ layer will be deposited on the top of the pSi grating. **b,** False-colored SEM image of a focused-ion beam cross section of a single unit cell for the fabricated sample after the RIE process. **c,** Cross-polarized transmittance spectrum of the metasurface at normal incidence illumination. The top panel depicts the cross-polarized transmission measurement setup. **d,** Measured optical dispersion for the GMR metasurface as obtained from the cross-polarized transmittance spectra and as a function of the incident angle. **e,** Measured diffraction efficiency (first-order reflection) of the metasurface. The inset shows the same data plotted on a logarithmic scale.

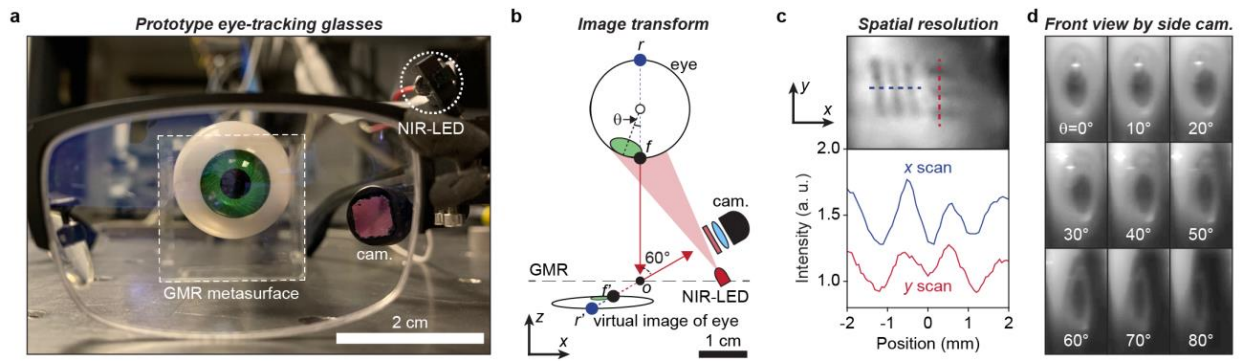

**Fig. 4 | Optical eye-tracking prototype demonstration. a,** Front image of the ET prototype showing the transparent high-$Q$ GMR metasurface in front of the eye. **b,** Schematic showing how the eye is imaged through a diffracted order of the metasurface. A ray-optics picture can be used to explain how the eye's image is compressed and pulled towards the metasurface. This action affords facile capture of large eye rotations. **c,** Imaging of 0.63-mm-wide vertical and horizontal bars from a USAF calibration chart by the ET prototype (a). The bottom panel shows cross section cuts of the intensity along horizontal (blue line) and vertical (red line) directions. **d,** Optical images taken from the eye with the ET camera. A wide range of rotation angles (up to $\theta = 80°$) can be imaged and the image resolution is sufficient to quantify the size and motion of the sclera, iris, and pupil.

# Supplementary Information

## S1. Optical constants for RCWA simulations

In order to simulate the behavior of the GMR metasurfaces, we use RCWA simulations which take into account the optical properties of the relevant materials. We set a constant value of the refractive index of $SiO_2$ (=1.45) to describe the quartz substrate and the capping layer. The refractive index spectra of LPCVD grown pSi and $Si_3N_4$ are experimentally characterized by spectroscopic ellipsometry. Figure S1a plots the complex refractive index spectrum of the pSi in the range from 500 nm – 900 nm. The real part of the refractive index ($n$) shows a gradual decrease from 4.5 to 3.7 while the imaginary refractive index ($k$) decreases from 0.2 to 0 as a function of the increasing wavelength. A peak is also seen near a wavelength of 800 nm in the $k$ spectrum. This is consistent with quantum confinement effects in the pSi grains or the presence of defects created during the LPCVD deposition[1,2], which could result in a blue-shifted bandgap[3,4]. This makes the peak diffraction efficiency for the TM GMR near 800 nm be smaller than that without the absorption. This will be discussed in more detail in section S5. Figure S1b shows that the magnitude of the refractive index of $Si_3N_4$ is virtually constant at 2.0 across the spectral region of interest.

## S2. Anti-reflecting layer design

Figure S2 plots the direct (zeroth-order) transmission spectra for the single-layer (200-nm-thick $Si_3N_4$) and the double-layer (100-nm-thick $SiO_2$/200-nm-thick $Si_3N_4$) antireflection coatings on the quartz substrate. The $Si_3N_4$ slab supports Fabry-Pérot transmission resonances at 800 nm and 400 nm, where the layer fits a full or half wavelength of light in the nitride material. The role of the $SiO_2$ capping layer is to further enhance the transmission in the spectral region where the $Si_3N_4$ slab could not. It displays a 290 nm optical path length difference (=2×1.45×100 nm) between the light reflected at the top and bottom interfaces of the capping layer. This corresponds to an approximately $\pi$ phase difference at the transmission dip of the single $Si_3N_4$ slab spectrum (~580 nm). The two layers combined offer a high transmission across the entire visible spectral range.

## S3. Grating coupling of free-space radiation to the GMR

The thin pSi grating is used to couple free-space radiation to the GMR and vice versa. Figure S3 shows the optical dispersion of the $TE_0$ (Fig. S3a) and $TM_0$ (Fig. S3b) waveguide modes for the stack of $SiO_2$/Si/$Si_3N_4$ layers on a quartz substrate. The dashed black, cyan, and yellow lines are representing the light lines of the air, the $SiO_2$, and $Si_3N_4$ media, respectively. The blue and red solid lines in Figs. S3a and S3b are the dispersion curves for the $TE_0$ and $TM_0$ waveguided modes, respectively. If a grating with a period $p$ is formed in the pSi layer, the incident planewave can couple to the quasi-guided modes by picking up a photon momentum equal to integer multiple of $2\pi/p$. The red horizontal arrow pointing to the right represents a second-order grating coupling process (i.e. a transfer of $4\pi/p$ photon momentum) from a normally-incident planewave to a guided mode. This corresponds to the excitation of the GMR. When photons are coupled into the waveguide, they propagate along the $Si_3N_4$ slab for $Q/2\pi$ optical cycles. Here, $Q$ is the *Q*-factor of GMR and the relation between the *Q*-factor and propagation time will be discussed in the later section (S13). In the near-infrared (NIR), the absorption is low and the guided light slowly leaks back into the 1st-order diffracted beams. This process is captured by the horizontal arrow pointing to the left. For a 1-μm-periodic grating, the in- and out-coupling phase matching conditions are satisfied at 870 nm for TE polarization, which well agrees with the spectral location of the TE GMR in the experiments (864 nm).

**S4. Temporal coupled-mode theory and power flow analysis of the GMR metasurface**

A quantitative study of the interactions between the GMR and the free-space radiation channels provides us with a deeper understanding of the power flow in the device. It also can be used to craft intuitive guidelines for the design of the GMR metasurfaces. Figure S4a describes the power flow, starting from the incident planewave (black arrow) to the output radiation channels (red arrows) as mediated by the GMR metasurface (grey arrows). Here, we focus on the GMR at 870 nm and TE polarization as an example.

We define the amplitudes of the incident planewave (*S*), the GMR (*a*), and the output radiation channels ($A^{(i)}_{sca,\sigma}$), where *i* is the diffraction-order number and *σ* is the radiation direction (i.e. transmitting or reflecting). By applying temporal coupled-mode theory[5], we build the following differential equation describing the temporal evolution of the amplitude *a* of the GMR:

$$\frac{da}{dt} = i\omega a = i\omega_0 a - \frac{1}{2}\left(\sum_{i,\sigma}\gamma^{(i)}_{sca,\sigma} + \gamma_{abs}\right)a + i\sqrt{\gamma^{(0)}_{sca,refl}}S, \tag{S1}$$

where $\omega$, $\omega_0$, $\gamma^{(i)}_{sca,\sigma}$, and $\gamma_{abs}$ are the angular frequency of the planewave, the resonant frequency of the GMR, the fractional decay rates of GMR into a specific radiation channels ($A^{(i)}_{sca,\sigma}$), and the non-radiative decay rate of GMR, respectively. The first, second, and third terms on the right-hand side of Eq. S1 describe the harmonic oscillation, the amplitude decay, and the excitation of the GMR, respectively. Using Eq. S1, we can obtain the relevant amplitudes and total decay rate for the GMR metasurface:

$$a = \frac{i\sqrt{\gamma^{(0)}_{sca,refl}}S}{i(\omega-\omega_0) + \frac{1}{2}\gamma_{tot}}, \tag{S2}$$

$$A^{(i)}_{sca,\sigma} = -\frac{\sqrt{\gamma^{(i)}_{sca,\sigma}\gamma^{(0)}_{sca,refl}}S}{i(\omega-\omega_0) + \frac{1}{2}\gamma_{tot}}, \tag{S3}$$

$$\gamma_{tot} = \sum_{i,\sigma}\gamma^{(i)}_{sca,\sigma} + \gamma_{abs}. \tag{S4}$$

The diffraction efficiency on resonance ($\omega = \omega_0$) can then be expressed as follows:

$$\eta^{(i)}_{sca,\sigma} = \frac{|A^{(i)}_{sca,\sigma}|^2}{|S|^2} = \frac{4\gamma^{(i)}_{sca,\sigma}\gamma^{(0)}_{sca,refl}}{\gamma_{tot}\gamma_{tot}} = 4f^{(i)}_{sca,\sigma}f^{(0)}_{sca,refl}, \tag{S5}$$

where $f^{(i)}_{sca,\sigma}$ is the fractional decay rate into an output radiation channel. The multiplicative form of the fractional decay rates on the right-hand side of Eq. S5 follows from the fact that the overall diffraction results from the consecutive processes of in-coupling from the incident planewave to the GMR ($f^{(0)}_{sca,refl}$) and the subsequent out-coupling of the guided light to a diffractive radiation channel ($f^{(i)}_{sca,\sigma}$).

By introducing absorptive materials into the GMR metasurface, we can control the resonance to be either radiative ($f_{abs} \ll 1$) or non-radiative ($f_{abs} \sim 1$). Control over the spectral absorption is the key mechanism by which rainbows are suppressed in the visible spectral range. Figure S4b shows the calculated power efficiencies into selected energy coupling channels. Here, the geometric parameters for the GMR metasurface are the same as for our experimental demonstration. The only difference is that we here sweep the imaginary part of the permittivity of the pSi while fixing its real part to $13.69\varepsilon_0$. In the simulations, we apply a monochromatic planewave at 870 nm from the top and calculate the power flux into each coupling channel. By normalizing the power flux to that of the incident planewave, we obtain the power efficiency: $\eta^{(i)}_{sca,\sigma}$ for the radiative channels. $\eta_{abs}$ is the power efficiency for the non-radiative decay channel. The extinction efficiency, $\eta_{ext}$ is the sum of all the efficiencies. For the vanishing imaginary part of the Si permittivity, $\eta_{abs}$ is zero, making the radiative decay dominate; $N$ number of the radiative channels almost equally contribute to the extinction ($f^{(i)}_{sca,\sigma} \sim 1/N$) and the numerically calculated diffraction efficiency into the first-order reflection (~11.1%) agrees well with that obtained by Eq. S5 with the presence of 6 radiative channels ($4 \times 1/6 \times 1/6 = 0.11$). In the NIR spectral region, the GMR is in the 'radiative regime' and we can perform optical eye-tracking (ET). The situation changes when the pSi has a non-vanishing imaginary permittivity. Due to the high $Q$-factor of the GMR, even a small amount of absorption of the pSi grating lets $\eta_{abs}$ dominate the total extinction efficiency ($\eta_{ext}$). The radiative decay process becomes extremely inefficient and the GMR enters a 'non-radiative regime.' At 605 nm, for instance, the imaginary permittivity of pSi is ~$0.8\varepsilon_0$ and this results in a radiative decay efficiency below $10^{-3}$. The non-radiative decay (i.e. absorption) prevents energy build up in the GMR and the extinction efficiency is effectively reduced as the magnitude of the imaginary part of the pSi permittivity increases. Figure S4c plots the fractional decay rate, where $f^{(i)}_{sca,\sigma}$ is defined as $\eta^{(i)}_{sca,\sigma}/\eta_{ext}$. This figure shows the non-radiative process can rapidly dominate the radiative processes as the imaginary part of the pSi permittivity is increased. When Im[$\varepsilon_{si}$] is $0.8\varepsilon_0$ which is the case for the pSi permittivity at 605 nm, the values, $f^{(0)}_{sca,refl}$ and $f^{(\pm 1)}_{sca,tran}$ become $1.5 \times 10^{-2}$ and $1.0 \times 10^{-2}$, respectively. Plugging in these values into the Eq. S5 predicts the diffraction efficiency ($\eta^{(\pm 1)}_{sca,tran}$) to be $6.0 \times 10^{-4}$. This quantitatively agrees well with our RCWA simulations ($\eta^{(\pm 1)}_{sca,tran} = 5.1 \times 10^{-4}$) shown in Fig. S4b, confirming the validity of our intuitive temporal coupled-mode theory.

The decreasing extinction efficiency as a function of the imaginary pSi permittivity is closely correlated to the increasing (decreasing) specular transmission (reflection), as can be seen in Fig. S4d. With lossless pSi metasurface elements in the near-infrared spectral region, the extinction efficiency is ~60% and the GMR redirects the input planewave into six diffractive orders. This makes the transmission (reflection) become lower (higher) compared to the case without the grating. In the visible spectral region, on the other hand, the extinction efficiency becomes lower than 4% and the resultant transmittance is higher than 80%, making the GMR a transparent, rainbow-free glass.

## S5. Transmission and diffraction efficiency spectra for the high-$Q$ GMR metasurface

Our high-$Q$ GMR metasurface manifests high transmission (> 80%) and acceptable diffraction (> 10%) as shown in Fig. S5. TE (TM) polarization are defined as having the electric fields parallel (perpendicular) to the groove direction, respectively. The broad features in the transmission spectra result from the Fabry-Pérot resonances of the dielectric stack. The narrow linewidth features are produced by the GMRs. The weak contrast in the refractive index between the $Si_3N_4$, the $SiO_2$, and the air makes the $Q$-factor of the Fabry-Pérot resonances low, resulting in broad transmission dips. Narrow, deep transmission dips in the NIR spectral region are indicative of the high-$Q$ of the GMRs. The dips in the visible spectral region are hardly noticeable due to the pSi absorption. This is consistent with the observed suppression of the rainbows. The lower diffraction efficiency (< 10%) of the TM GMR (~800 nm) than that of the TE polarization is due to the pSi absorption in this spectral range, as discussed in section S1.

## S6. Transition of GMR from nonradiative to radiative regime

Figure S6 plots the fractional decay rate into absorption (= $f_{abs}$) in the NIR ($\lambda$ = 870 nm) for the GMR metasurface as a function of grating height. We take a representative refractive index of pSi (=3.7+0.005i) in this spectral regime. For a small grating height (< 2 nm), the radiative decay of GMR is weak, causing the non-radiative decay to dominate (>50%) the total decay. When the grating height reaches 2 nm, the fractional absorption falls below 50% and the peak diffraction efficiency exceeds 4%.

## S7. Fabrication of GMR structures

Figure S7 summarizes the GMR metasurface fabrication process. A quartz substrate is prepared (Fig. S7a) by a standard RCA cleaning: 1) Removing insoluble organic contaminants with a $H_2O:H_2O_2:NH_4OH$ (5:1:1) solution for 10 minutes at 50°C. 2) Removing ionic and heavy metal atomic contaminants using a $H_2O:H_2O_2:HCl$ solution (6:1:1) for 10 minutes at 50°C. A 200-nm-thick $Si_3N_4$ film is deposited by a low pressure chemical vapor deposition (LPCVD) at 800°C (Fig. S7b). A 50-nm-thick pSi is then deposited by the same technique at 620°C (Fig. S7c). Wet oxidation at 950°C gradually converts the pSi into $SiO_2$, starting at the top. In our GMR metasurface we leave a 3-nm-thick Si layer on the $Si_3N_4$ film (Fig. S7d). Photolithography patterns are fabricated onto a 2 cm × 2 cm area (Fig. S7e) and RIE etching transfers the lithographic grating patterns into the pSi layer (Fig. S7f). We remove the remaining photoresists and the oxidized $SiO_2$ grooves by consecutively putting the sample in a piranha solution and a buffered oxide etch (BOE) solution (Fig. S7g). A silica capping layer is finally fabricated by spin-coating hydrogen silsesquiaxane (HSQ, 4% in methyl isobutyl ketone) and curing at 400°C for 12 hours (Fig. S7h).

## S8. Precise control over the pSi grating height by wet oxidation

The slow rate of the wet oxidation process at 950°C is ideally suited to control the pSi layer thickness with atomic-scale precision. Figure S8 plots the thickness of pSi as a function of oxidation time. In order to characterize the thickness, we prepare several sheets of the bare Si wafers and deposit 200-nm-thick $Si_3N_4$ film and 50-nm-thick pSi. By varying the oxidation time, we obtain wafers with different pSi thickness which we characterize using spectroscopic ellipsometry. pSi with a 2.8 nm thickness is obtained after a 30 minutes oxidation time.

## S9. Effect of over-etching of the pSi grating elements

As shown in Fig. 3b, our GMR structure features a 30-nm-deep over-etching into the $Si_3N_4$ slab. This increases the grating coupling strength and lowers the radiation $Q$-factor, as can be seen in Fig. S9. Here, we calculate the first-order reflection by RCWA simulations which take into account over-etching depths ranging from 10 nm to 40 nm. Both the TE and TM GMR peaks show an increasing bandwidth in addition to a slight blue-shift as a function of the over-etching depth. For 30-nm-thick over-etching, the $Q$ factor is calculated to be 1400. The background non-resonant diffraction also shows a monotonic increase as a function of the over-etching depth.

We note a different behavior for the peak diffraction efficiency of the TE GMR at 870 nm. Its peak value hardly changes while the other peak in the visible spectral range show an increase as a function of increasing over-etching depth. The TE mode at 608 nm, TM mode at 800 nm, and TM mode at 580 nm all exhibit a growing peak diffraction efficiency. The invariant peak diffraction efficiency at 870 nm for the TE mode is due to the negligible absorption at this wavelength in pSi, which is not seen for the other wavelengths. According to the coupled-mode theory developed in the previous section (S4), the peak diffraction efficiency is expressed as $4f^{(i)}_{sca,\sigma} f^{(0)}_{sca,refl}$, where the value of the fractional radiative decay rate, $f^{(i)}_{sca,\sigma}$ is competing with that of the absorption. In the absence of absorption, the fractional radiative decay rate hardly changes when the grating is over-etched. In the presence of (even very weak) absorption, the process of over-etching increases the radiative decay rate, and thus also the peak diffraction efficiency.

## S10. Angle-resolved spectroscopy to map the GMR dispersion properties

In order to characterize the optical dispersion as well as the strength of the GMRs, we constructed a home-built angle-resolved spectroscopy setup (Fig. S10). This setup can be switched between a 'cross-polarized transmission mode (Fig. S10a)' and a 'diffraction mode (Fig. S10b)'. In the cross-polarized transmission mode, the broadband light source is fixed to the optical axis and only the GMR metasurface is rotated with respect to the optical axis. The incident broadband planewave passes through a polarizer which is oriented at 45° with respect to the Si strips of the GMR metasurface. The analyzer is oriented at 90° with respect to the polarizer and the light can pass through it only when it interacts with the GMR metasurface. The spectra are normalized by the incident planewave spectrum. The measured spectra as a function of incident angle directly map the optical dispersion of both the TE and TM modes (Fig. 3d). In diffraction mode, the broadband light source illuminated the GMR metasurface at normal incidence. The sample and source are simultaneously rotated with respect to the collection optical axis. Depending on the frequency, the diffracted light fluxes will be focused on different locations at the back focal plane of the collection lens. A 1-mm-wide aperture stop at the back focal plane of the collection lens enables us to select the frequency of the diffracted light collected into the detection system. By rotating the samples, the collected light frequency is swept and the diffraction-efficiency spectrum is obtained. The effective numerical aperture (NA) is 1/400, which corresponds to a collection area of approximately 1 mm².

**S11. Image formation with the GMR metasurface**

We develop a coordinate transformation rule to predict the image location by a diffractive ray-tracing method. Different from specular reflection images, the diffracted image is located near the plane of the GMR metasurface. This enables the ET camera located along the diffraction angle ($\theta_o$) to image the eye, regardless of the distance between the eye and the GMR plane. This is explained in more detail in the section S12. Figure S11 depicts the transformation from an object point ($P$) to the corresponding diffractive ray image point ($P'$). Here, we assume the GMR mode is positioned at the $\Gamma$-point ($\theta = 0°$) such that the normally-incident planewave would excite the GMR in optimal condition, as is the same case for the experiments. One ray from the point $P$ in red is normally-incident on the GMR metasurface while the other ray (blue) is incident at an angle, $\theta_i$. The normally incident ray intersects with the GMR surface at the origin, $O(x_o, 0)$ and diffracted across an angle $\theta_o$. The diffracted ray can be expressed in terms of the coordinate system ($x_i$, $z_i$) in the following way:

$$z_i = \cot\theta_o \cdot (x_i - x_o). \tag{S6}$$

Another ray (blue) with an incident angle $\theta_i$ strikes at ($x_o+z_o\tan\theta_i$, 0) and the diffracted ray can also be expressed as follows:

$$z_i = \cot\theta_d \cdot (x_i - x_o - z_o \tan\theta_i), \tag{S7}$$

where $\theta_d$ is the diffraction angle for the incident angle of $\theta_i$. Combining Eq. S6 and Ep. S7, we obtain a linear set of equations:

$$\begin{bmatrix} x_i \\ z_i \end{bmatrix} = \begin{bmatrix} 1 & -\dfrac{\tan\theta_i \tan\theta_o}{\tan\theta_d - \tan\theta_o} \\ 0 & -\dfrac{\tan\theta_i}{\tan\theta_d - \tan\theta_o} \end{bmatrix} \cdot \begin{bmatrix} x_o \\ z_o \end{bmatrix}. \tag{S8}$$

Equation S8 lets us pinpoint the location of image points $P'(x_i, z_i)$ for different locations on the object $P(x_0, z_0)$.

When the incident angle ($\theta_i$) is larger than $\theta_c$ (grey arrow in Fig. S11), the incident planewave hardly excites the GMR as it is outside of the GMR spatial bandwidth. Details about $\theta_c$ will be discussed in the following section (S13). For a $Q$-factor of 1400 for example, $\theta_c$ is $\pm 0.25°$. The small value of $\theta_c$ allows us to approximate Eq. S8 in an even simpler, more compact form:

$$\begin{bmatrix} x_i \\ z_i \end{bmatrix} = \begin{bmatrix} 1 & -\cos^2\theta_o \sin\theta_o \\ 0 & -\cos^2\theta_o \cos\theta_o \end{bmatrix} \cdot \begin{bmatrix} x_o \\ z_o \end{bmatrix}. \tag{S9}$$

The first column of the transformation matrix shows that a vector component parallel to the GMR plane is transformed as the same vector. The length is also conserved. The interesting properties of the diffracted image are related to the second column. It suggests that a surface-normal vector is rotated over an angle $\theta_o$ toward the diffracted ray output while its magnitude is reduced by a factor of $\cos^2\theta_o$. If a camera is oriented toward the diffracted ray output, it would capture the front view of the object even when located at an oblique direction. At the same time, the diffracted image is squeezed along the $z$-direction, virtually attached to the GMR surface. This makes it always possible to capture the eye, no matter how far it is located from the GMR surface. This is not achievable by imaging through a conventional reflector.

### S12. Diffractive ray imaging affords a large depth of field

Another technological advantage of diffractive ray imaging with the proposed metasurface is the large depth of field. The surface-normal vector in the object space does not only change the orientation, but also shrinks in its magnitude by the factor of $\cos^2\theta_o$ (Eq. S9), where $\theta_o$ is the diffraction angle (see the transformation of a displacement vector **V** to **V'** in Fig. S12). This locates the diffractive image of the eye closer to the GMR surface (by an amount of $\cos^3\theta_o$ as compared to that of the eye in the object space), as calculated in Fig. S12. The distance between the GMR surface and the eye in black (red) is 3 cm (9 cm). The corresponding diffractive eye image is 0.375 cm (1.125 cm) from the GMR surface. This would be easily captured by the ET camera if it is optimized to focus on the GMR surface. In order to demonstrate this, we take the diffractive images of an artificial eye by locating it at 3 cm and 9 cm from the GMR surface. We observe clear eye images in both cases, which would be difficult using conventional reflectors (especially for 9 cm). The slightly increased blurriness of the image for 9 cm is due to the finite depth of field of the ET camera, which can be improved.

### S13. $Q$-factor of GMR and spatial resolution of diffractive ray imaging

Figure S13a depicts the energy flow process from the incident planewave to the diffracted output far-field radiation channels. Due to the presence of the surface relief grating on the nitride waveguide, the incident planewave can be coupled to a GMR (Fig. S3). The quasi-guided waves then propagate along the $x$-direction inside the $Si_3N_4$ slab waveguide until they are decoupled to create the diffracted beams via the grating out-coupling. The propagation length ($L$) for the quasi-guided waves can be expressed as follows:

$$L = v_g \tau = v_g Q / \omega_r = \frac{\lambda_r}{n_g} \cdot \frac{Q}{2\pi},$$ (S10)

where $v_g$, $\tau$, $Q$, $\omega_r$, $\lambda_r$ and $n_g$ are the group velocity, total decay time, $Q$-factor, angular resonant frequency, resonant wavelength in the air, and group index of the relevant waveguide mode, respectively. The right-hand side of Eq. S10 provides an intuitive picture where $L$ is proportional to the $Q$-factor of the GMR. This length is proportional to the number $Q/2\pi$ of optical cycles and the wavelength $\lambda_r/n_g$ of the quasi-guided modes.

The finite propagation length $L$ in the $x$-direction directly set the bandwidth of available spatial frequencies in the $x$-direction as:

$$\Delta k_x = \frac{2\pi}{L} = k_r \cdot \frac{2\pi n_g}{Q}.$$ (S11)

This spread in spatial frequency requires that the magnitude of $k_x$ of the incident planewave to be smaller than $\Delta k_x$ in order to excite the GMR (Fig. S13b). By taking the ratio between $\Delta k_x$ and the magnitude of $\mathbf{k}^{(0)}$ ($k_r$), the maximum acceptable angle of incidence ($\theta_c$) is obtained as $\sin^{-1}\Delta k_x/2k_r$. Fig. S11 reminds us that among the rays from the point $P$, only the rays within a $\theta_c$ range are able to excite the GMR and be efficiently diffracted. This determines the angular components we can collect from the object ($P$) through the GMR and sets the fundamental limit on the effective NA ($\sin\theta_c = \pi n_g/Q$). In order to characterize the resolving power, we take the diffractive ray images of the discs with different diameters (Fig. S13c, d). We are able to resolve and this is sufficient to distinguish the relevant components of human eyes. For more resolving power, we need to enlarge the acceptance angle by lowering the $Q$-factor and make wider angular components to be resonantly diffracted.

### S14. Functional glasses for Augmented Reality applications

To highlight the benefits of our technology, we summarize a comparative overview of the form factor (compactness), transparency, severity of rainbow issues, ease of mass production, optical efficiency, and computational load of existing technologies[6] in table S1. Approaches that employ conventional optics (e.g. bulk mirrors, beam-splitters, and prisms) to redirect light reflected from the eye towards an imaging system can achieve a high-transparency in the visible without notable rainbow issues. However, the achievable optical power efficiency for the imaging process is unavoidably constrained as this quantity is traded off with the transparency. Their practical applicability is also severely limited due to their bulkiness and ponderousness especially for wearable, light-weight AR devices. Diffractive optical elements can resolve the issues of bulkiness and the constrained optical power efficiency of the imaging system. These elements include periodically patterned gratings, high-index polymers, holographic gratings, and liquid crystals on the few-micrometer scale. A number of companies pursue the use of such components. Through coherent scattering (diffraction), the diffractive optical elements achieve high optical power efficiency, even when composed of transparent materials (e.g. glass). Their structural flatness greatly reduces the volume and bulkiness. Also, the fabrication process is relatively simple, which is advantageous for mass production with top-down lithography processes. However, the presence of a grating is inevitably accompanied by undesired diffraction effects in the visible spectral range – referred to as the rainbow issue. Some approaches engage heavily darkened glasses (absorptive) to reduce the rainbow issues at the cost of a strongly reduced transparency.

     Our approach based on a high-$Q$ GMR addresses all outstanding issues: bulkiness, manufacture complexity, optical power efficiency, and challenges with rainbows. It is comprised of a few-nm-thick Si grating on a half-$\lambda$ thin dielectric film, which further enhances its compactness and light-weight. Due to the superb elastic properties of $Si_3N_4$, our proposed GMR structures are applicable even to curved surfaces. The most significant attribute of the high-$Q$ GMR is the fact that all the wave front manipulation (e.g. the diffraction process) is mediated by the high-$Q$ resonances. As a result, all diffraction phenomena can be turned off when the high-$Q$ resonance is damped, even in the presence of only minute optical absorption rates. This allows even few-nm-thick, spectrally selective absorptive materials to turn on and off all the beam redirecting processes without lowering the transparency of the overall system.

As such, the users' vision is barely perturbed by undesired diffraction (rainbow issues), even in the bright day light (see Fig. 1d). We also fabricate the high-$Q$ GMR structures with standard Si-compatible processes, such as optical lithography and top-down dry etching techniques, which are compatible with mass production schemes. As discussed in the following supplementary information section (S16), we also numerically demonstrate the potential for more advanced asymmetric GMR designs that offer even higher diffraction efficiency (~53%) by full-field RCWA simulations.

### S15. Demonstration of accurate object tracking using diffracted images

To directly demonstrate the tracking ability of our proposed diffractive imaging method, we show that a one-to-one correspondence can be obtained between the actual angular locations of a target object (e.g. an eye) and the measured positions. Figure S14a describes the experimental setup that is used for this purpose. A NIR-LED illuminates a 4-mm-wide circular target object which represents the eye; In real life it could correspond to the pupil or the boundary between the iris and sclera. The light scattered from the target object towards the GMR surface undergoes a diffractive redirection and is collected by the ET camera system. The CMOS sensor in the ET camera system captures the diffracted image of the circular target object (Fig. S14b). For further analysis, we convert this image into a binary image (Fig. S14c) using a simple threshold selection method and extract the $x$ and $y$ coordinates of its centroid position (red crosses in Fig. S14b,c) with Matlab's built-in fitting functions. We mount the target object on two different rotation stages which rotate the polar and azimuthal angle, respectively. Due to a slight displacement of the target center from the polar axis ($\delta_p$) and the displacement of the azimuthal axis from the polar axis ($\delta_a$), the $x$ and $y$ coordinates of the target object vary as being subject to the polar and azimuthal angle rotations (Fig. S14d). There is a finite distance between the target (e.g pupil) and the center of the eye ($=\delta_p$). In our experiment, the polar and azimuthal rotation axes are also not perfectly crossing and there is a small offset ($=\delta_a$). By taking into account these displacements, we express the $x$ and $y$ coordinates of the centroid of the target object as a set of closed functional forms:

$$x = (\delta_p \sin\theta_o - \delta_a) \cdot \sin(\varphi - \varphi_o) \tag{S12}$$

$$y = \delta_p \{\cos\theta_o - \cos(\theta + \theta_o)\} \cdot \sin(\varphi - \varphi_o) \tag{S13}$$

where $\delta_p$, $\delta_a$, $\theta_o$, and $\varphi_o$ are the displacement of the target center from the polar axis, the displacement of the azimuthal axis from the polar axis, and the initial polar (azimuthal) angle when the polar (azimuthal) angle rotation stage is set to zero, respectively. Figure S14e-h present the measured $x$ (Fig. S14e) and $y$ (Fig. S14g) coordinates and their modelled coordinates (Fig. S14f,h) obtained from Eqs. S12, S13. By accurately setting the displacement parameters ($\delta_p = 3.3$ mm, $\delta_a = 2.5$ mm, $\theta_o = 105°$, and $\varphi_o = 6°$), the modelled coordinates match well to the experimentally measured coordinates. Deviations at $\theta = 20°$ and $25°$ are due to the non-ideal illumination conditions which hinder the extraction of the binary image of the target.

The presence of a mathematically closed form allows us to transform the ($x$, $y$) coordinate of the target object to the corresponding polar and azimuthal angles:

$$\theta = \cos^{-1}\left(\frac{\delta_p \cos\theta_o - y}{\delta_p}\right) - \theta_o \tag{S14}$$

$$\varphi = \sin^{-1}\left(\frac{x}{(\delta_p \sin\theta_o - \delta_a)}\right) + \varphi_o . \tag{S15}$$

Figure S15a,b compare the measured azimuthal and polar angles with the actual angles which are set to the rotation stages. All the measured data show excellent agreement with the actual angles within ~1° deviation, consistent with what is generally required in commercial eye tracking systems. The aberrations that arise by imaging through a diffracted-order have been studied extensively[7,8]. Typically, they are quite limited and distortion dominates. A distortion would lead to an observed shape change of the circular target beyond the expected shape change that result from looking at the target from a grazing angle (See Fig. S12). To analyze the possible impact of distortion, we observe the target while performing an azimuthal rotation. For the measurements, we set the polar angle to be 0°. In Fig. S15c, we plot the measured diameter of the target object along the horizontal direction (minor axis length). Providing that the ET camera is located at 60° from the surface-normal direction of the GMR, the diameter of the circular target object along the horizontal axis is reduced by the amount of cos(60°) (= 0.5) when the azimuthal angle is set to the initial condition ($\varphi$-$\varphi_o$=0). The initial minor axis length is measured to be 1.89 mm, which is ~0.5 of the target object diameter. The 0.11 mm deviation is attributable to the binary image conversion process and the spatial resolution of our GMR along the horizontal axis. Under the rotation of the azimuthal axis from -50° to 50° the measured minor axis length quite closely follows the expected cosine behavior. Only at large positive and negative angles the impact of distortion is visible. Such distortions can be corrected for in a calibration and do not take away from the ability to track the motion of the eye. The findings in Fig. S15 demonstrate that our proposed diffractive ray imaging method can also be applied to measure the size of a pupil. Another opportunity we expect from this is the calibration of GMR surface-normal gazing direction. Considering that the minor axis length reaches its maximum value when the azimuthal angle is matched to the surface-normal direction, we are able to identify when the eye is oriented to the surface-normal direction by monitoring the minor axis length of pupil or iris. In the present case, we have used this method to determine the initial azimuthal angle, $\varphi_o = 6°$.

**S16. Guided-mode resonance designs for asymmetric diffraction**

One of the insights we obtain from the coupled-mode analysis is that the diffraction efficiency into a desired diffraction output channel can be increased by suppressing the fractional decay rate into the other, non-desired output channels (Eq. S5). As such, it would be advantageous to 'activate' only one output diffractive order and suppress the others to enhance the power efficiency. In this section, we introduce two different approaches to achieve highly-asymmetric diffraction properties and, based on these, we numerically demonstrate >50% asymmetric diffraction performances through full-field RCWA simulations.

The first approach is by manipulating the interference between the polarization currents induced in the 3-nm-thick Si gratings. When a *y*-polarized planewave illuminates the GMR structure at near-normal incident angle (0.5°), a fraction of the energy couples into the GMR mode (See Fig. S16a). The high-*Q* GMR the guides the coupled light inside the Si$_3$N$_4$ slab for $Q/2\pi$ optical cycles. This establishes a well-defined electric field distribution on the top surface of the Si$_3$N$_4$ slab, where the 3-nm-thick Si grating elements are located. Given the very thin (3 nm) grating height, we regard the grating as a weak out-coupler and describe all the diffraction output phenomena in the framework of perturbation theory. Application of the first-order Born approximation[9] provides us the complex coupling amplitude (Huygens amplitude) into the ±*m*$^{th}$ diffractive order as the following expression:

$$I(\pm m) = -i\omega\varepsilon_o \int_{-p/2}^{p/2} \chi(x) E_y(x) e^{-i\frac{2\pi x}{\lambda}\sin\theta} e^{\mp i\frac{2m\pi x}{p}} dx .  \quad (S16)$$

The Huygens amplitude into the ±*m*$^{th}$ diffractive order, $I(\pm m)$ is defined as a definitive integration over a unit cell on the top surface of the Si$_3$N$_4$ slab where the 3-nm-thick Si gratings will be located. ±*m*, $\omega$, *p*, $\varepsilon_o$, $\chi(x)$, $\lambda$, and $\theta$ are the output diffractive-order number, angular frequency of the light, grating period, vacuum permittivity, relative susceptibility distribution function, vacuum wavelength of light, and the incident angle, respectively. Equation S16 describes that the mutual interference between the infinitesimal polarization currents ($-i\omega\mathbf{P}=-i\omega\varepsilon_o\chi\mathbf{E}$) at each location *x* weighted by a kernel wave function of the ±*m*$^{th}$ diffractive order determines the Huygens amplitude. For simplicity, we will omit the proportionality constant, $-i\omega\varepsilon_o$ in the following discussions.

We further simplify Eq. S16 by combining the electric field and the zeroth-order phase factor ($e^{-i2\pi x\sin\theta/\lambda}$):

$$I(\pm m) = \int_{-p/2}^{p/2} \chi(x) f(x) e^{\mp i\frac{2m\pi x}{p}} dx = \int_{-p/2}^{p/2} j(x) \left\{ \cos\left(\frac{2\pi mx}{p}\right) \mp i\sin\left(\frac{2\pi mx}{p}\right) \right\} dx \quad (S17)$$

, where *f(x)* and *j(x)* represent the combined electric field $\{=E_y(x)e^{-i2\pi x\sin\theta/\lambda}\}$ and the polarization current density $\{=\chi(x)f(x)\}$, respectively. If we decompose the polarization current density into even and odd components with respect to the central plane of the unit cell (*x*=0), we obtain an even simpler expression:

$$I(\pm m) = \int_{-p/2}^{p/2} j_e(x)\cos\left(\frac{2\pi mx}{p}\right) dx \pm \int_{-p/2}^{p/2} -i\cdot j_o(x)\sin\left(\frac{2\pi mx}{p}\right) dx = I_e(m) \pm I_o(m) . \quad (S18)$$

We define $I_e(m)$ and $I_o(m)$ as the even and odd contributions to the Huygens amplitude. The even and odd components of the polarization current density $\{j_e(x)$ and $j_o(x)\}$ can be simply obtained by the following operations:

$$j_e(x) = \frac{1}{2}\{j(x) + j(-x)\} \quad (S19)$$

$$j_o(x) = \frac{1}{2}\{j(x) - j(-x)\} . \quad (S20)$$

Note that the contribution of the even polarization current to the Huygens amplitude is invariant under the sign change of diffractive order while that of the odd polarization current inverts its sign. As such, we can accomplish a large magnitude contrast between the Huygens amplitudes between the +*m*$^{th}$ and -*m*$^{th}$ diffractive orders if we find the optimum Si grating locations that makes the $I_e(m)$ and $I_o(m)$ almost equal; $I(+m)$ $\{=I_e(m)+I_o(m)\}$ obtains a large amplitude while $I(-m)$ $\{=I_e(m)-I_o(m)\}$ vanishes.

The investigation of the local electric field distribution at the top surface provides the key information to find the optimum grating locations for high contrast $\pm m^{th}$ order Huygens amplitudes. In Fig. S16b, we plot the even $\{h_e(x)\}$ and odd $\{h_o(x)\}$ integrand functions obtained from the full-field RCWA simulations:

$$h_e(x) = f_e(x)\cos\left(\frac{2\pi mx}{p}\right) \tag{S21}$$

$$h_o(x) = -i \cdot f_o(x)\sin\left(\frac{2\pi mx}{p}\right). \tag{S22}$$

The integrand functions, $h_e(x)$ and $h_o(x)$ mean the even and odd local Huygens amplitude from a pair of locations ($x$, $-x$) occupied by infinitesimal grating element of unit ($=1$) susceptibility. Due to the $-i$ factor at the right-hand side of Eq. S22, $f_e(x)$ and $f_o(x)$ are required to have varying phase profiles as a function of $x$; otherwise, $I_e(m)$ and $I_o(m)$ would have a $\pi/2$ phase difference, resulting in the same magnitude of $I(+m)$ and $I(-m)$. The choice of exact normal incidence, in this regard, is not preferable for obtaining effective asymmetric diffraction. It produces standing wave where the phase profile is invariant. Allocating grating elements of complex-valued susceptibility could be an option to achieve the asymmetric diffractions upon the normal incidence. However, we will pursue another option by utilizing our GMR mode without resorting to lossy materials. As originating from a waveguide mode in the $Si_3N_4$ slab, the high-$Q$ GMR mode has a rapid phase propagation along the $x$ direction. In our case, the propagation constant ($\beta_x$) is $4\pi/p+2\pi/\lambda\times\sin(\theta_i)$ where $p$, $\lambda$, and $\theta_i$ are the grating period, the wavelength of light, and the incidence angle, respectively. The most dominant term, $4\pi/p$ is the second-order grating momentum the incident light gains when coupled into the GMR mode. The propagating nature of GMR allows the combined field components, $f_e(x)$ and $f_o(x)$ to have varying phase profiles as a function of $x$ as well as their mutual phase difference even at a near-normal incident angle ($\theta_i=0.5°$). As can be seen in Fig. S16b, we found two matching points ($x=\pm112.5$ nm) where $h_e(x)$ and $h_o(x)$ are equal. These locations are advantageous for achieving highly-asymmetric diffraction if activated as polarization current sources.

Integration of the $h_e(x)$ and $h_o(x)$ functions as multiplied by the susceptibility distribution of the grating, $\chi(x)$ gives the even and odd Huygens amplitudes, $I_e(m)$ and $I_o(m)$. Here, we focus on two regions ($-175\leq x \leq-50$, $50 \leq x \leq 175$ nm, as indicated as red-shaded regions in Fig. S16b) around the two matching points over which the definite integrals of the $h_e(x)$ and $h_o(x)$ functions are almost the same. We locate two Si elements (per period) at these locations and let only the polarization currents in these two ranges contribute to the integrations; the susceptibility function, $\chi(x)$ vanishes in the absence of the Si. Figure S16c plots the complex Huygens amplitudes integrated over the two regions. Due to the similarity of $I_e(m)$ and $I_o(m)$ both in magnitude and phase, the Huygens amplitude for the $+m^{th}$ diffractive order $\{I_e(m)+I_o(m)\}$ obtains a large magnitude while that of the $-m^{th}$ diffractive order $\{I_e(m)-I_o(m)\}$ is almost vanishing. We apply this Si grating location to the full-field RCWA simulations and observe a 53% peak diffraction efficiency into the $+1^{st}$ order at the GMR frequency ($=907.3$ nm) as plotted in Fig. S16d. We find that the extinction ratio of the magnitude squared Huygens amplitudes between the two diffractive orders (=148.06) matches quite well to the extinction ratio of the peak diffraction efficiencies (=148.12) between the two diffractive orders obtained by the RCWA simulations with only 0.04% error, which also confirms the validity of our perturbation analysis. The local electric field intensity distribution in Fig. S16e clearly shows the onset of nearly-asymmetric diffraction. We also remark on the robustness of the electric field distribution of the GMR mode upon the change of the incident angle or $\chi(x)$. The dominant contribution of the grating momentum ($4\pi/p$) to the propagation constant of the local electric fields ($\beta_x$) makes the location of the two matching points less sensitive to the change of the incident angle. This allows the highly asymmetric diffraction conditions to continue even for 7° incident angle. This condition also holds for several tens of nanometer errors either for the position and the width of the Si grating elements.

The second approach is by asymmetric waveguiding as depicted in Fig. S17a where a $y$-polarized plane wave is incident on the GMR structure with a small (6°) angle of incidence. Here, the non-vanishing $x$ component of the 0$^{th}$ order $k$-vector, $k_x^{(0)}=2\pi\sin\theta/\lambda$ makes the +1$^{st}$ diffractive order be guided into the Si$_3$N$_4$ waveguide while the -1$^{st}$ diffractive order still remain as free-space radiation. This situation holds when the incident angle, $\theta$ satisfies the following condition:

$$\theta > \sin^{-1}\left(1-\frac{\lambda}{p}\right) \tag{S23}$$

, which requires the GMR wavelength to approach the grating period in order to achieve asymmetric diffraction with a small angle of incidence. To shift the GMR frequency to a longer wavelength compared to that of our ET demonstration (870 nm), we make the Si$_3$N$_4$ slab slightly thicker (300 nm), which allows the GMR to appear at 946 nm for a 6° incident angle. As Figs. S17b,c show, the diffractive reflection occurs only into the -1$^{st}$ diffractive order and this results in the enhanced peak diffraction efficiency into the -1$^{st}$ order above 40% (Fig. S17d). This asymmetric waveguiding condition would hold as long as the incident angle is larger than 2.8° according to Eq. S23.

**S17. Independent control of spectrally-decoupled optical functions in wavelength-division multiplexing**
A unique attribute of the proposed high-$Q$, non-local metasurfaces is a wavelength multiplexing capability. Given the high-$Q$ of the GMRs, it is possible to multiplex or de-multiplex optical beams that have closely spaced frequencies and narrow spectral bandwidths. It allows one to integrate different optical functions (e.g. beam redirection, beam splitting, and polarization control) by simply stacking a number of functional high-$Q$ metasurface layers operating at different wavelengths. The introduction of a spectrally-selective absorbing materials to each layer then allows one to individually control (i.e. turn on or off) the particular optical function of each layer without affecting the function of the other layers. To clarify this idea, we numerically investigate a multiplexed system consisting of three different, stacked, high-$Q$ metasurfaces that provide independent control over the optical beams with closely spaced wavelengths.

The schematics in Figs. S18a-c describe how three flat optical elements can be stacked to allow multiplexing of optical beams at three closely-spaced wavelengths. As the initial step, we prepare three high-index layers ($n = 2.0$) that can support quasi-guided modes and afford good anti-reflection behavior (i.e. a high overall transmission). These layers are comprised of a 100-nm-thick $Al_2O_3$ ($n$ = 1.7) capping layer and a $Si_3N_4$ slab. The thicknesses of $Si_3N_4$ slabs at the top, middle, and bottom hetero slab are 90 nm, 220 nm, and 200 nm, respectively to achieve good antireflection behavior. The high-index layers are separated by 707-nm-thick $SiO_2$ spacer layers. The overall device provides a high transmissivity across a broad spectral range, as shown by the dashed curves in Fig. S18d. Next, we introduce a 3-nm-thick surface relief grating ($n$ = 3.5) into the top high-index layer to turn it into a guided-mode resonator (Fig. S18a). The period of the grating was chosen to be 847 nm, which enables the resonant excitation of the GMR with a normally incident beam with a 710 nm wavelength. The transmission dip and reflection peak in the solid curves in Fig. S18d highlight the efficient reflection of light on resonance. The high-$Q$ factor (=12,000) ensures a spectrally narrow bandwidth (0.06 nm) for this optical function; the flow of light outside this narrow spectral region is hardly affected. In a following step, we can sequentially also turn the middle and bottom layers into GMRs by inserting 3-nm-amplitude and 770 nm period surface relief gratings into these layers (Figs. S18b,c). The addition of the middle and the bottom high-$Q$ GMR modes lets the combined three-layer stack structure interface also with the normally incident beams at 705 nm (middle), and 700 nm (bottom), respectively (Figs. S18e,f). Importantly, the introduction of a GMR in one of the layers does not impact the response at the other wavelengths and this allows a simple stacking and adding of functions. Such behavior is not seen for low-$Q$ metasurfaces, where there is typically a very strong optical interaction between stacked layers.

      The chosen operation wavelengths and their spectral separations are also freely reconfigurable by modifying the geometric parameters (i.e. grating period, grating height, slab thickness, filling fraction of gratings, and materials). The three optical functions at 710 nm, 705 nm, and 700 nm are not only spectrally decoupled (Fig. S18f) but also spatially separated as can be seen in the local electric field intensity profiles of Figs. S18g-i. This allows us to also demonstrate the independent control of the three optical functions by allocating three different absorbing materials to each of the grating layers.

      Figure S19 summarizes our investigation of the independent control over the optical functions of the three high-$Q$ GMR metasurface layers. To show that their independent control is physically feasible, we include plausible Lorentzian oscillators into the 3-nm-thick gratings of the three high-$Q$ GMR metasurface layers. The resonance wavelengths of the Lorentzian oscillators for the top, middle, and bottom gratings match the resonant wavelengths of the top, middle, and bottom high-$Q$ GMR modes, respectively. We set the linewidth of the oscillators to be 10 nm, which is the same as the reported exciton linewidth of a monolayer $WS_2$ in the room temperature[10]. We also set the oscillation amplitude to be 1/20 of the reported exciton amplitude of a monolayer $WS_2$. Regarding the successful demonstrations of exciton resonance tuning via solid-state[11] and ionic liquid[12] gating, it is reasonable to assume that the amplitude of these oscillators can completely be turned on and off by injecting free charge carriers. Figure S19a shows the simulated transmissivity and reflectivity spectra when the top grating oscillator is turned on (grey curve in Fig. S19a) while the other oscillators are off. The oscillator at the top grating generates optical absorption, which significantly dampens the high-$Q$ GMR mode in the top metasurface. As a result, the transmissivity and reflectivity at

the resonant wavelength (710 nm) are almost restored as is the case for the absence of the top high-$Q$ GMR metasurface layer. This shows one of the unique traits of our proposed spectrally-decoupled high-$Q$ metasurfaces. The optical absorption in the thin (e.g. 3 nm) gratings can not only deactivate the optical functions themselves but also retrieve the flow of light as if there were no optical element present. In our experiments, we utilize the natural optical absorption of pSi for erasing all the beam redirection functions in the visible spectral range to suppress the rainbow effect.

Due to the spatial and spectral separation between each of the metasurface layers, the operations of the middle and the bottom high-$Q$ GMR metasurface layers are hardly affected. Their resonance locations and linewidths remain the same. Even ~1.5% changes in the depth of the transmission dips or the height of the reflection peaks are due to the modification of the background transmissivity and reflectivity spectra from the adjusted refractive index of the 3-nm-thick top grating layer. If the linewidth of the Lorentzian oscillators were narrower, the modification of the background transmissivity and reflectivity would be even smaller. We also investigate the independent control of the middle (Fig. S19b) and the bottom (Fig. S19c) high-$Q$ GMR metasurface layers, in which the optical behaviors are the same. We expect our investigation can evolve into more elaborate, practical designs for efficient, optoelectronic devices in which metasurfaces can conveniently be stacked and activated on demand.

**Supplementary references**

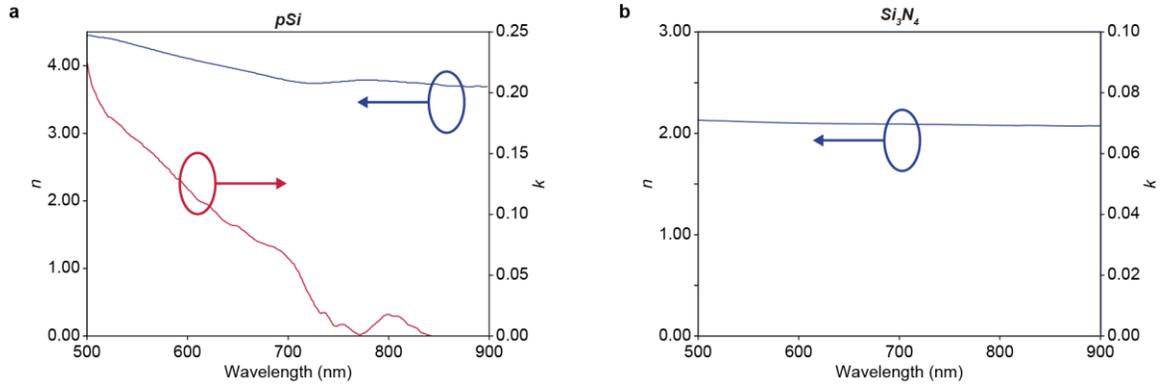

**Fig. S1 | Complex refractive index of grown films. a,** Refractive index spectra of LPCVD pSi film. Blue and red curves are representing the real and imaginary part of the refractive index, respectively. **b,** Refractive index spectra of LPCVD $Si_3N_4$ film.

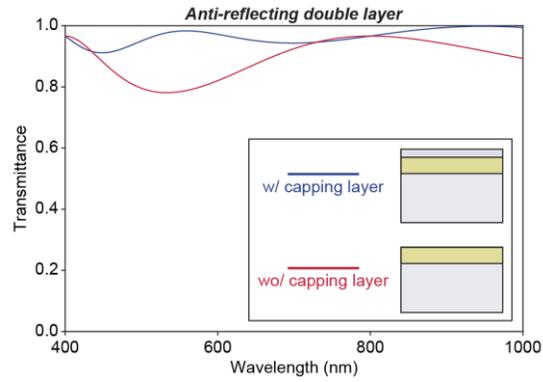

**Fig. S2 | Anti-reflecting double layer design.** The blue and red curves are the transmittance spectra from $SiO_2/Si_3N_4$ double-layer and a $Si_3N_4$ single-layer antireflection coatings, respectively. The thicknesses of the $SiO_2$ and $Si_3N_4$ layers are 100 and 200 nm, respectively. The spectra are calculated by RCWA simulations under the normal planewave incidence from the top.

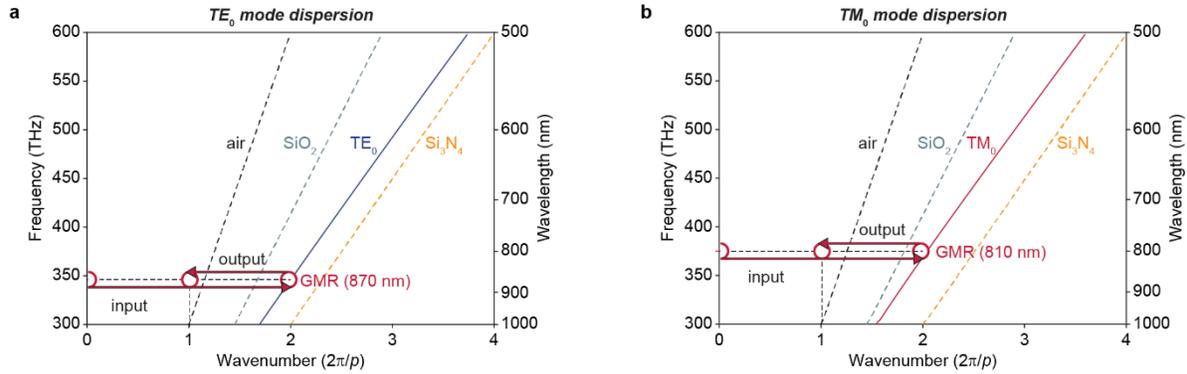

**Fig. S3 | Grating coupling mechanism for GMR excitation. a,** Optical dispersion curve of the $TE_0$ waveguide mode in the $Si_3N_4$ slab. This is calculated by the transfer-matrix method for an air/100-nm-thick $SiO_2$ slab/3-nm-thick Si film/200-nm-thick $Si_3N_4$ slab/quartz stack. The black, cyan, and yellow dashed curves are the light lines for air, $SiO_2$, and $Si_3N_4$, respectively. The horizontal red arrow to the right direction indicates the in-coupling process for a normally-incident planewave to the $TE_0$ waveguide mode, which corresponds to the excitation of the GMR. The red arrow to the left direction represents the out-coupling process from the $TE_0$ waveguide mode to free-space radiation into the first diffracted order. **b,** Optical dispersion curve for the $TM_0$ waveguide with the same structure as in (a).

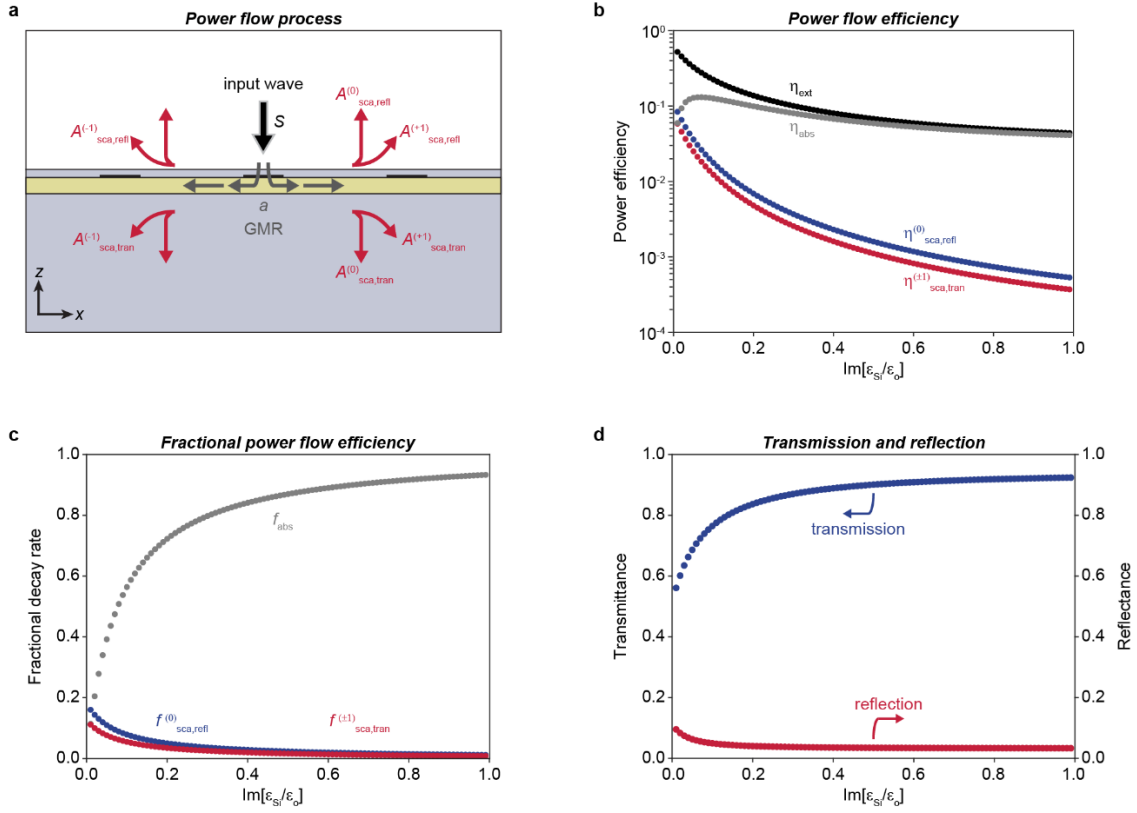

**Fig. S4 | Power flow analysis on the GMR metasurface. a,** Schematic defines all relevant out-coupling radiation channels and describes their contribution to the power flow. $S$, $a$, and $A^{(i)}$ are the amplitude of incident planewave, GMR, and the $i^{th}$ diffracted order, respectively. **b,** Power efficiency into each out-coupling channel as a function of the imaginary part of the Si permittivity. **c,** Selected fractional decay rate for each out-coupling channel. **d,** Zeroth-order transmittance and reflectance at the GMR resonance frequency (870 nm) as a function of the imaginary Si permittivity.

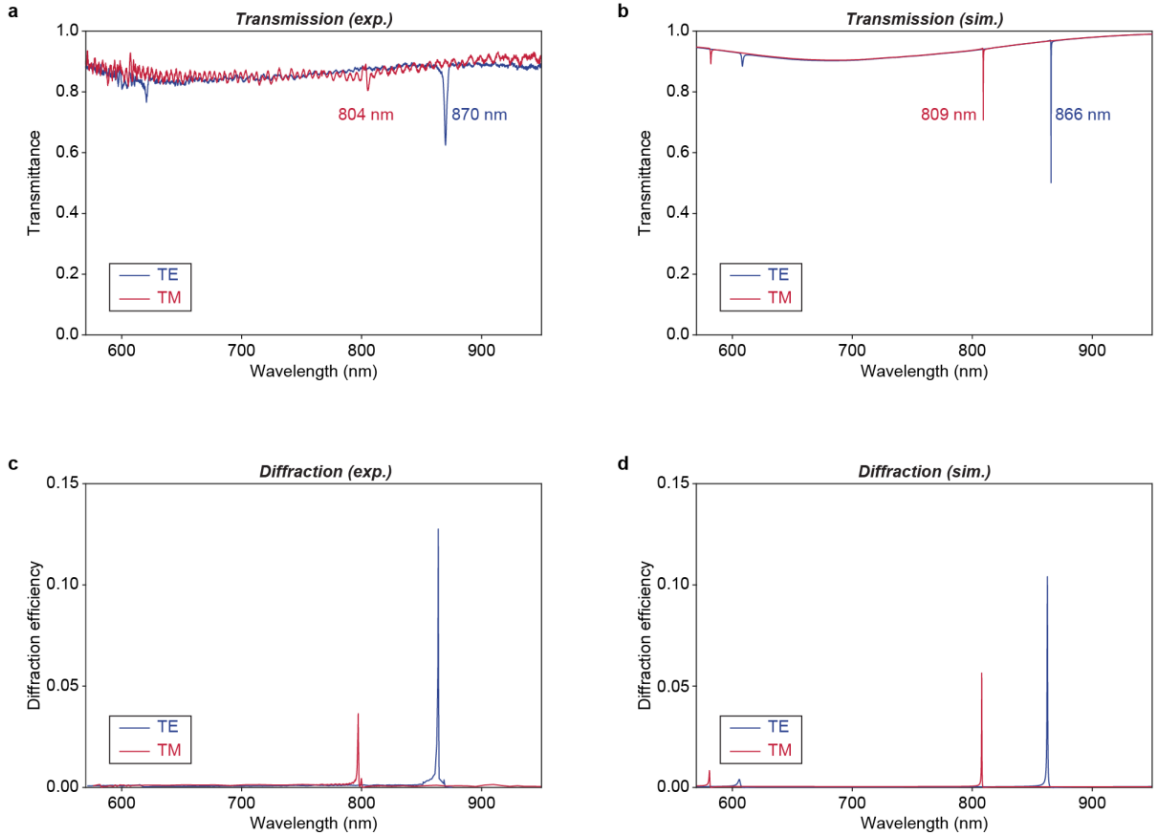

**Fig. S5 | Comparison of experimental and simulated spectra. a,** Measured specular transmission spectra for the TE and the TM polarizations. **b,** Simulated specular transmission spectra obtained with RCWA simulations. **c,** Measured diffraction efficiency spectra. **d,** Simulated diffraction efficiency spectra.

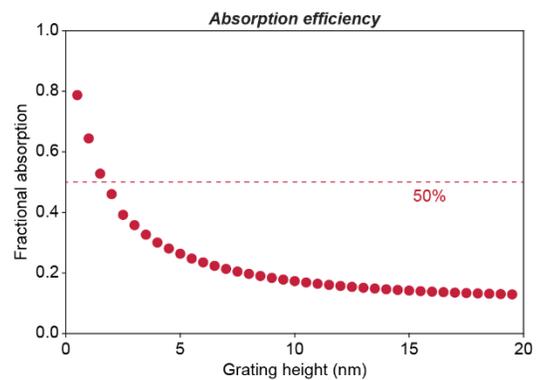

**Fig. S6 | Fractional absorption efficiency for the GMR in NIR spectral region.** Simulated fractional decay rate into absorption in pSi for TE GMR at 870 nm.

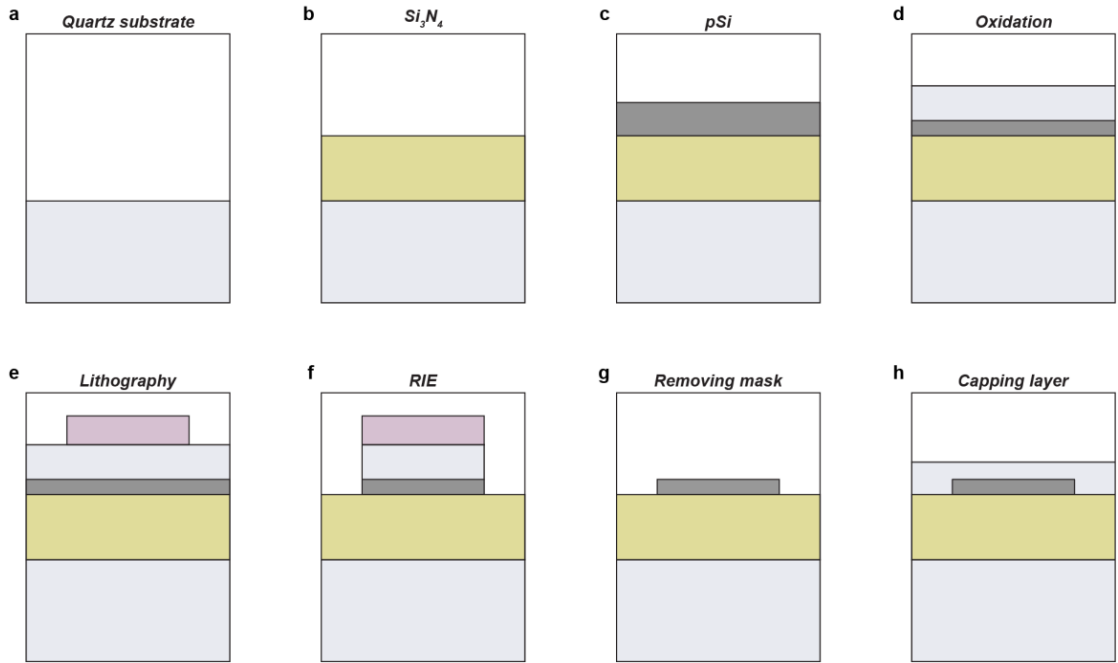

**Fig. S7 | Fabrication of GMR structures. a,** Preparation of quartz substrate. **b,** 200-nm-thick $Si_3N_4$ deposition with LPCVD. **c,** 50-nm-thick pSi deposition with LPCVD. **d,** Wet oxidation until 3-nm-thick pSi remains. **e,** Dry etching mask formation by photolithography. **f,** RIE etching. **g,** Removal of etching mask by wet etching with buffered oxide etch (BOE). **h,** 100-nm-thick $SiO_2$ capping layer coating with HSQ.

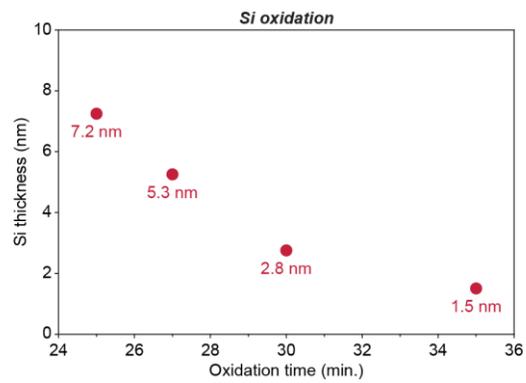

**Fig. S8 | Si thickness control by wet oxidation.** pSi thickness as a function of wet oxidation time at a temperature of 950°C. The initial pSi thickness is 50 nm (at 0 minutes).

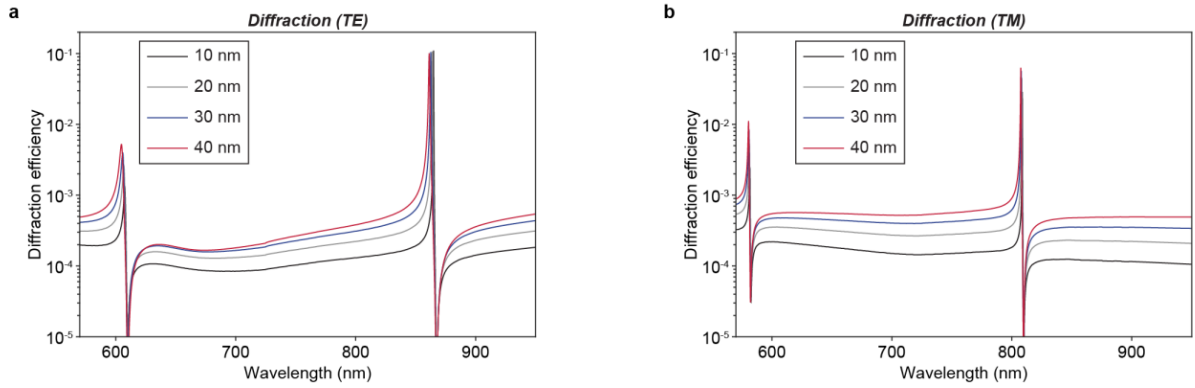

**Fig. S9 | Effect of over-etching. a,** Diffraction efficiency spectra for TE GMR as a function of the over-etching depth into $Si_3N_4$ slab. **b,** Diffraction efficiency spectra for TM GMR as a function of the over-etching depth into $Si_3N_4$ slab.

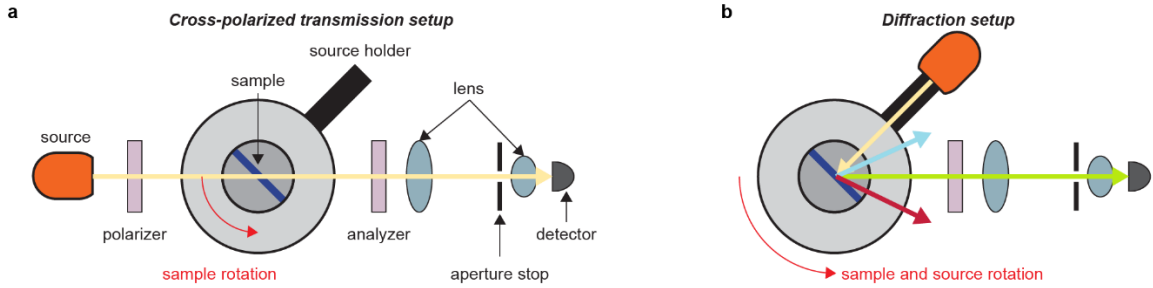

**Fig. S10 | Angle-resolved spectroscopy setup. a,** Angle-resolved spectroscopy setup in cross-polarized transmission mode. Broadband light source is fixed to the optical axis and the sample is rotated with respect to the optic axis. **b,** Angle-resolved spectroscopy setup in diffraction mode. The orientation of broadband light source is fixed to the sample, which is rotating with respect to the fixed optical axis.

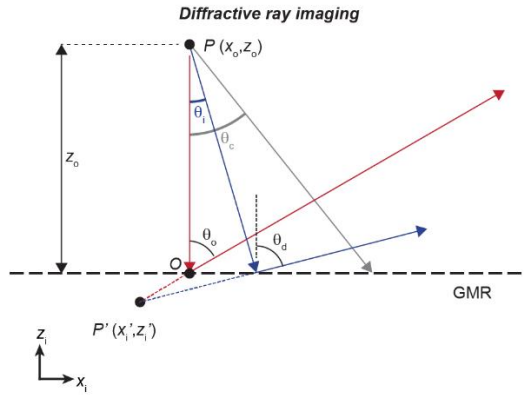

**Fig. S11 | Diffractive ray tracing transformation.** Mapping from an object point ($P$) to the corresponding virtual image point ($P'$) by diffractive ray tracing. The red arrow indicates the normally incident planewave which is diffracted into $\theta_o$. The blue arrow shows an obliquely incident planewave at the angle $\theta_i$ which is diffracted into $\theta_d$. The intersection of the two extrapolated dashed lines indicates the location of the virtual image ($P'$). ($x_i'$, $z_i'$) is the coordinate of the virtual image via diffractive ray tracing.

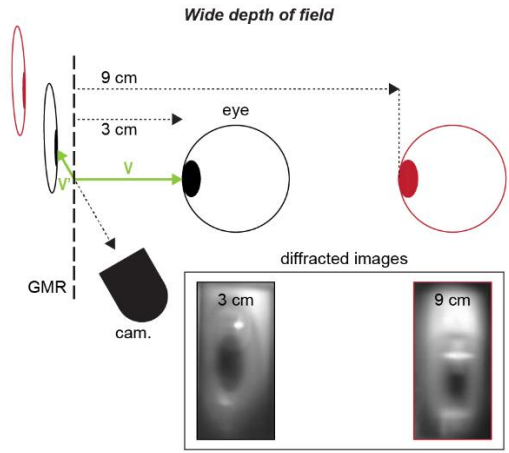

**Fig. S12 | Large effective depth of field.** Mapping of eyes with different distance from the GMR surface by applying Eq. S9. The inset shows the diffractive images of an artificial eye at different location from the GMR surface. **V** is a displacement vector from the GMR surface to a point of the artificial eye. It transforms into **V'** through diffractive ray imaging (Eq. S9).

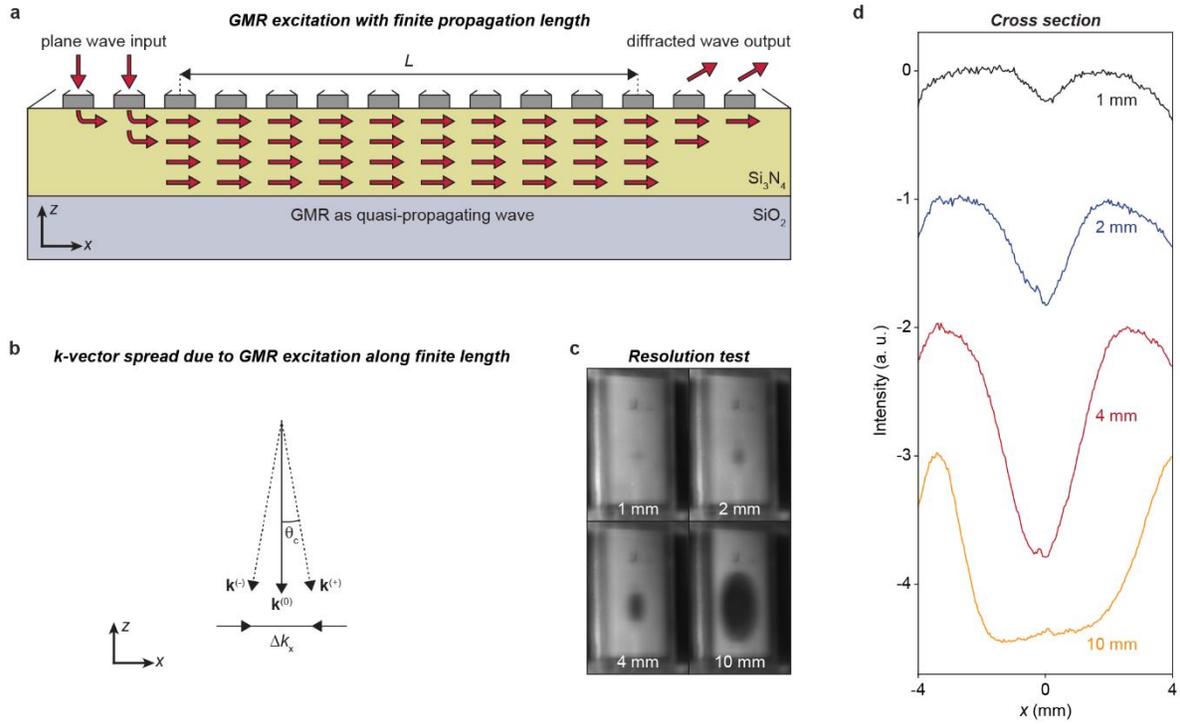

**Fig. S13 | Relation between *Q*-factor of GMR and spatial resolution of diffractive image. a,** Power flow diagram from the incident planewave to the diffracted output waves via the GMR. Due to the presence of the grating, the quasi-guided waves radiate to free space. *L* is the characteristic propagation length of the quasi-guided waves. **b,** Representation of **k**-vector range of the incident plane waves for GMR excitation. **c,** Spatial resolution test for diffractive ray images. Black discs with different diameters are imaged. The differences between 1 mm, 2 mm, 4 mm, and 10 mm discs are resolvable. This also suggests that the pupil size can be distinguished by our imaging system. **d,** Horizontal cross section of the diffracted images shown in (c). The labels indicate the diameter of individual disk and the cross sections are shifted by an offset for clarity.

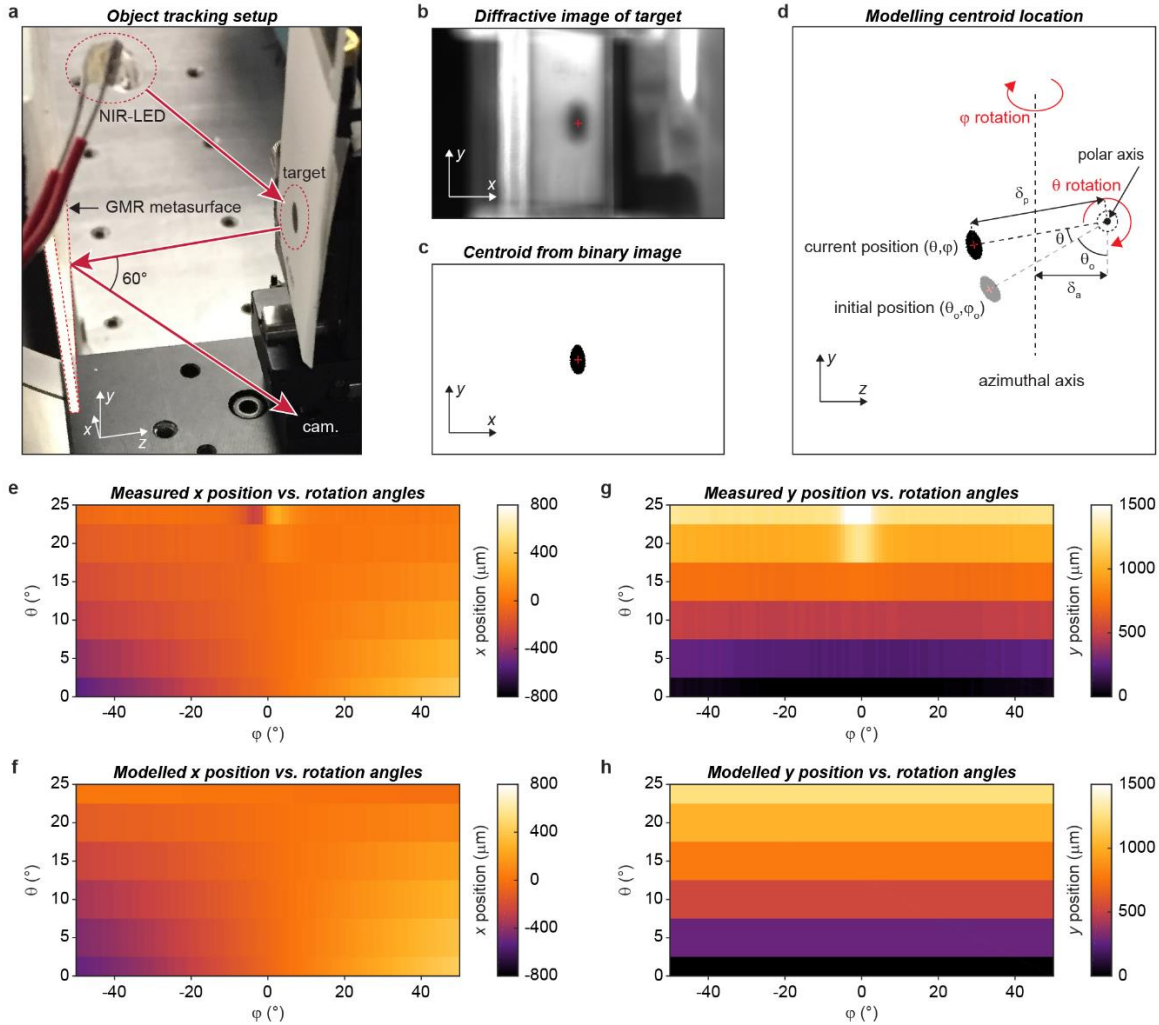

**Fig. S14 | Object tracking with our diffractive ray imaging method. a,** Experimental setup for object tracking demonstration. A NIR-LED illuminates a 4-mm-wide black circle target object. The scattered light by the target is incident on the GMR surface and diffracted into the 1$^{st}$ order reflection into the ET camera. To mimic the eye motion, the target object is mounted on two rotation stages; one for the polar angle rotation and another for the azimuthal angle rotation. **b,** Diffractive ray image captured by CMOS sensor of ET camera system. The red cross is the centroid of the black circle target which is determined from its binary image (c). **c,** Binary image of the target object extracted from the diffractive image (b). **d,** Schematic description of the displacement length of the center of the target object from the polar axis ($\delta_p$), the displacement length of the azimuthal axis from the polar axis ($\delta_a$), the polar angle ($\theta$), and the azimuthal angle ($\varphi$). **(e-h),** Measured $x$ (e) and $y$ (g) coordinates of the centroid of the target object and the modelled $x$ (f) and $y$ (h) coordinates of the centroid of the target object according to Eq. S12 and Eq. S13.

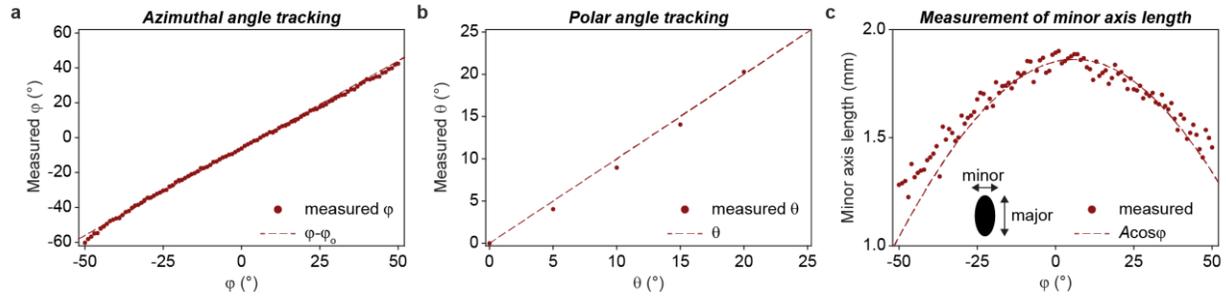

**Fig. S15 | Precise angle measurement. a,** Measured azimuthal angle of the target object using the closed mathematical formula in Eq. S15. The dashed red curve indicates the azimuthal angle location of the rotation stage. The polar angle of the rotation stage is set to be 0°. **b,** Measured polar angle of the target object using the closed mathematical formula in Eq. S14. The red dashed line indicates the polar angle location of the rotation stage. The azimuthal angle of the rotation stage is set to be $\varphi_o$ (=6°). **c,** Measured length of the minor axis along the horizontal direction as a function of azimuthal angle. The dashed red curve indicates the cosine curve with a fitted amplitude, $A$. The polar angle of the rotation stage is set to be 0°.

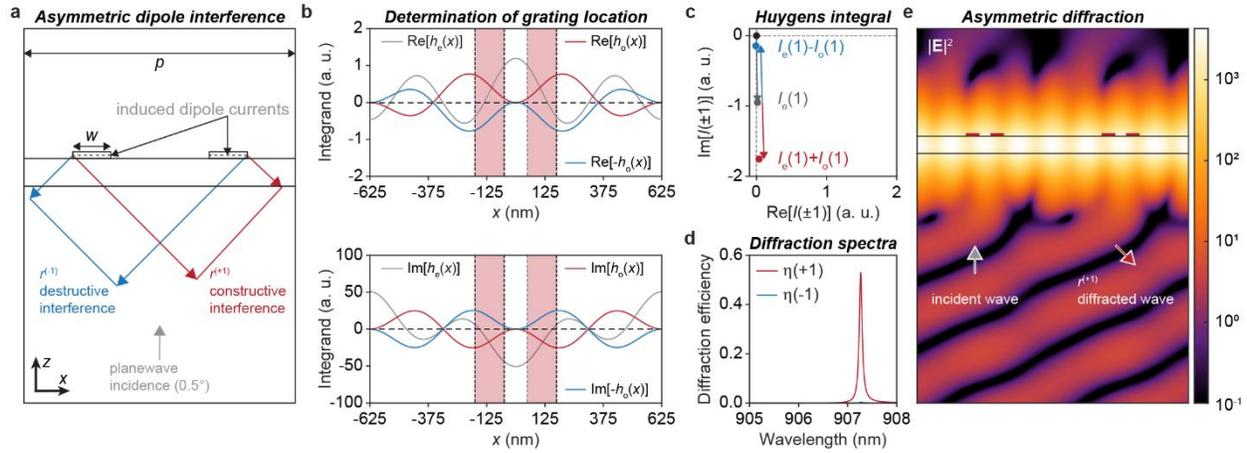

**Fig. S16 | Asymmetric GMR diffraction by interference of polarization current. a,** Schematic description of the asymmetric interference of polarization currents into the +1st (red) and -1st (blue) diffractive orders. The dipoles in the Si grating elements constructively interfere into the +1st diffractive order while destructively interfering into the -1st diffractive order. A unit cell is presented and the grating height, period (=$p$), and $Si_3N_4$ thickness are 3 nm, 1250 nm, and 138 nm, respectively. For simplicity we set the super- and substrate as vacuum. **b,** Even and odd components of the integrand function as a function of the $x$ coordinate on the top surface of $Si_3N_4$ slab. Grey, red, and blue solid lines are the even component into ±1st diffractive order, the odd component into +1st diffractive order, and the odd component into -1st diffractive order, respectively. The top (bottom) panel is the plot of the real (imaginary) part of each function. The two red-shaded regions indicate the range of $x$ which give similar definitive integral values of $I_e(1)$ and $I_o(1)$. **c,** Huygens amplitudes for the +1st diffractive order (red circle) and the -1st diffractive order (blue circle) in complex plane. **d,** Calculated diffraction efficiency spectra into the +1st (red) and the -1st (blue) diffractive orders by full-field RCWA simulations. **e,** Local electric field intensity distribution over two unit cells at the GMR frequency.

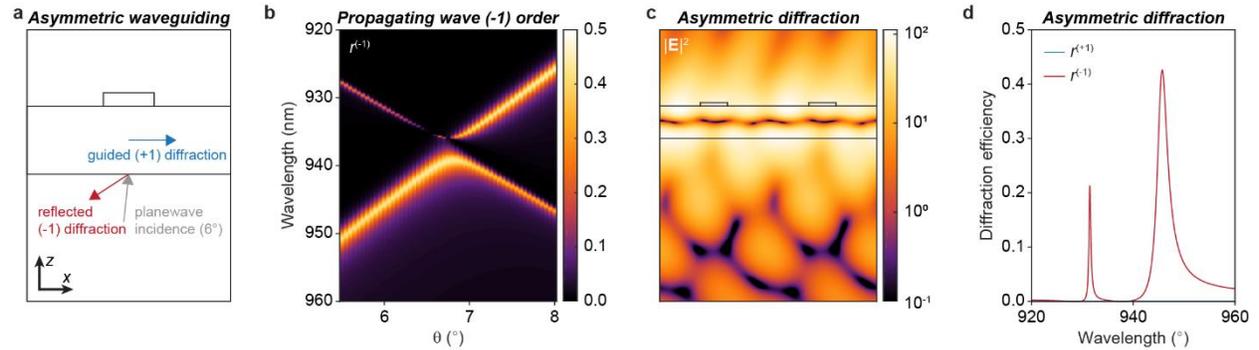

**Fig. S17 | Asymmetric waveguiding GMR structure. a,** Schematic description of a unit cell where the +1$^{st}$ diffractive order is guided while the -1$^{st}$ diffractive order is reflected into the free-space. The Si grating height is 3 nm and the over etched depth into the $Si_3N_4$ slab is 30 nm, similar to the fabricated GMR structure. The period, grating width, and $Si_3N_4$ slab thickness are 1000 nm, 250 nm, and 300 nm, respectively. For simplicity, we set the superstrate and substrate as vacuum. **b,** Calculated diffraction efficiency dispersion plot into the -1$^{st}$ diffractive order. **c,** Local electric field intensity distribution over two unit cells at the GMR frequency. **d,** Diffraction efficiency spectra into the -1$^{st}$ (red) and the +1$^{st}$ (blue) diffractive orders.

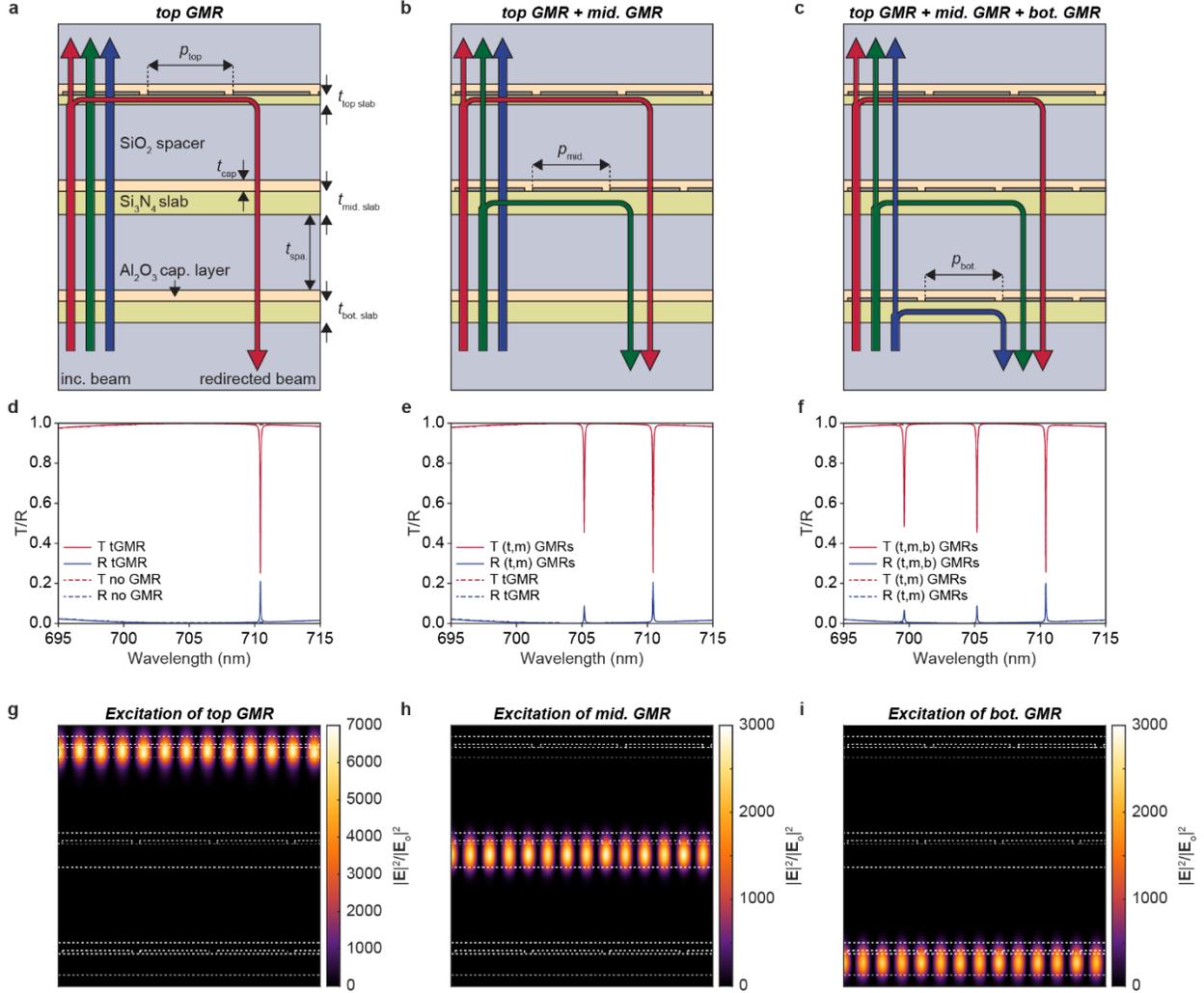

**Fig. S18 | Wavelength-division multiplexing of diffractive optical functions by stacking high-$Q$ GMR metasurfaces. a,** Cross section of the stack of three high-index layers. The introduction of 3-nm-thick surface relief gratings into each layer turns them into high-$Q$ GMR metasurfaces. $p_{top}$, $t_{top,\,slab}$, $t_{mid.,\,slab}$, $t_{bot.,\,slab}$, $t_{spa}$, and $t_{cap}$ are the period of the top grating, thickness of the top $Si_3N_4$ slab, thickness of the middle $Si_3N_4$ slab, thickness of the bottom $Si_3N_4$ slab, thickness of the $SiO_2$ spacer layer, and thickness of the $Al_2O_3$ capping layer, respectively. The three vertical arrows in red, green, and blue indicate the normally incident planewave at 710 nm, 705 nm, and 700 nm, respectively. The polarization direction is in parallel to the groove direction. The high-$Q$ GMR mode at the top high-index layer only interacts with the incident planewave at 710 nm, stores notable energy, and ultimately redirects the light into a reflected beam. At the same time, it allows free passage of the incident planewaves with wavelength of 705 nm and 700 nm. The filling fraction of the grating is 90%. **b,** Functionalization of the middle high-$Q$ GMR metasurface by adding a surface relief grating to the middle high-index layer. $p_{mid.}$ is period of the middle grating. The middle high-$Q$ GMR metasurface selectively interfaces with the incident planewave at 705 nm while allowing free passage of light at other wavelengths. **c,** Functionalization of the bottom high-$Q$ GMR metasurface by adding a grating on the bottom hetero slab. $p_{bot.}$ is the period of the bottom grating. The bottom high-$Q$ GMR metasurface selectively interfaces with the incident planewave at 700 nm. **d,** Specular transmission (red) and reflection (blue) spectra of the structure in (a). The dashed curves are the same spectra in the absence of any surface relief gratings. **e,** Transmission and reflection spectra of the structure in (b). The dashed curves are the same spectra of the structure (a) for comparison. **f,**

Transmission and reflection spectra of the structure in (c). The dashed curves are the same spectra of the structure (b) for comparison. **(g-i),** The local electric field intensity enhancement of the structure in (c) under a monochromatic planewave incidence at 710 nm (g), 705 nm (h), and 700 nm (i) wavelengths.

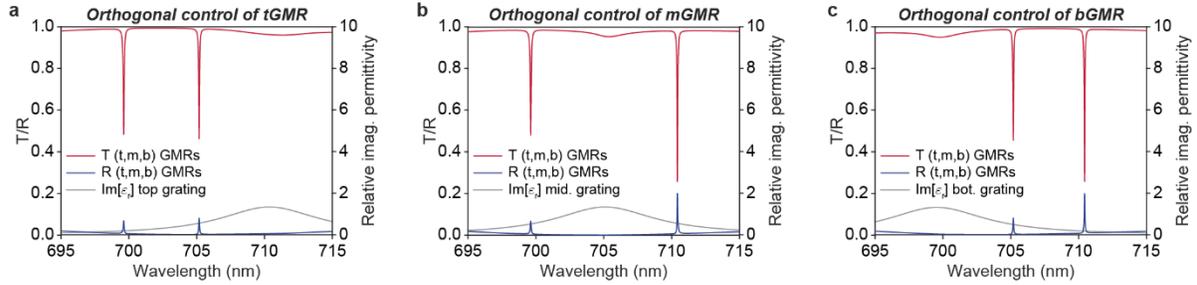

**Fig. S19 | Independent control of optical functions. a,** Specular transmission (red) and reflection (blue) spectra for the stack of three high-$Q$ GMR metasurfaces in Fig. S18c when the Lorentzian oscillator in the top high-$Q$ GMR metasurface is in its absorbing, 'on' state. The grey curve plots the imaginary part of relative permittivity of the dielectric grating in the top high-$Q$ GMR metasurface. **b,** Same spectra when the Lorentzian oscillator is placed in the middle high-$Q$ GMR metasurface is in its on-state. The grey curve plots the imaginary part of relative permittivity of the dielectric grating in the middle high-$Q$ GMR metasurface. **c,** Same spectra when the Lorentzian oscillator in the bottom high-$Q$ GMR metasurface is in its on-state. The grey curve plots the imaginary part of relative permittivity of the dielectric grating in the bottom high-$Q$ GMR metasurface.

**Table S1 | Functional wearable glasses for Augmented Reality**

|  | Conventional optics | Cameras and Image processing | Diffractive optics I | Diffractive optics II | High-$Q$ GMR |
|---|---|---|---|---|---|
| **Essential elements** | Bulk prisms or mirrors | Multiple cameras | Surface-relief gratings | Holographic gratings | Guided-mode resonators |
| **Representative companies/groups** | Google[19], Lingxi[20], Lochn[21], Lumus[22], NVIDIA[23,24], Optivent[25], Fuchs et al.[26], Hua et al.[27–30] | Argus Science[31], Ergoneers[32], EyeLink[33], Pupil Labs[34], SMI[35], Tobii[36] | Dispelix[37], Magic Leap[38], Microsoft[39,40], Waveoptics[41], | Apple[42*], Digilens[43], Microsoft[44,45], Sony[46], TruLife Optics[47], Lee et al.[48–50], Liu et al.[51], Li et al.[52] | Current work |
| **Structural attributes** | | | | | |
| Form factor | Bulky (cm) | Bulky (mm) | Compact (μm) | Compact (μm) | Compact (nm) |
| Manufacturing complexity | Complex assembly process | Complex assembly process | Mass production compatible | Mass production compatible | Mass production compatible |
| **Optical attributes** | | | | | |
| Transparency | High | High | Low | High (~80%) | High (~90%) |
| Rainbow issue | No dispersion | Absent | Severe | Severe | Low (< 0.1%) |
| Power efficiency | Low** | High | High | High | High*** |
| View | Front-view | Side-view | Front-view | Front-view | Front-view |
| **Electronic attributes** | | | | | |
| Computational load | Light | Heavy | Light | Light | Light |

* Apple recently acquired the eye tracking company Akonia.
** <10%
*** GMR designs with 53% diffraction efficiency have been demonstrated (Supplementary information S16).